\documentclass[final]{agujournal}

\journalname{Earth and Space Science}

\usepackage{adjustbox}

\begin{document}

%
%

\title{Benchmarking Geant4 for simulating galactic cosmic ray interactions within planetary bodies}

%
%

\authors{K.E. Mesick\affil{1}, W.C. Feldman\affil{2}, D.D.S. Coupland\affil{1}
 and L.C. Stonehill\affil{1}}

\affiliation{1}{Los Alamos National Laboratory, Los Alamos, NM 87545 USA}
\affiliation{2}{Planetary Science Institute, Tuscon, AZ 85719 USA}

\correspondingauthor{K.E. Mesick}{kmesick@lanl.gov}

\begin{keypoints}
\item Neutron and gamma-ray spectroscopy of planetary bodies provide information about hydrogen abundance and geochemical composition.
\item Radiation transport simulations of GCR interactions with these bodies are a key tool to predict and interpret data from such measurements.
\item In this paper we benchmark Geant4 simulations with the Apollo 17 Lunar Neutron Probe Experiment data.
\end{keypoints}

%
%

\begin{abstract}
Galactic cosmic rays undergo complex nuclear interactions with nuclei within planetary bodies that have little to no atmosphere.  Radiation transport simulations are a key tool used in understanding the neutron and gamma-ray albedo coming from these interactions and tracing these signals back to geochemical composition of the target.  We study the validity of the code Geant4 for simulating such interactions by comparing simulation results to data from the Apollo 17 Lunar Neutron Probe Experiment.  Different assumptions regarding the physics are explored to demonstrate how these impact the Geant4 simulation results.  In general, all of the Geant4 results over-predict the data, however, certain physics lists perform better than others.  In addition, we show that results from the radiation transport code MCNP6 are similar to those obtained using Geant4.
\end{abstract}

%
%

\section{Introduction}

Monte Carlo based radiation transport simulations have become a standard technique for understanding systems in a wide range of applications.  Of relevance to this paper, the application of radiation transport codes to studying the interaction of galactic cosmic rays (GCRs) within planetary surfaces has emerged as a robust technique for predicting and interpreting measurements from neutron and gamma-ray spectroscopy instruments.  

\begin{figure}[h!]
\centering
\includegraphics[width=0.7\textwidth]{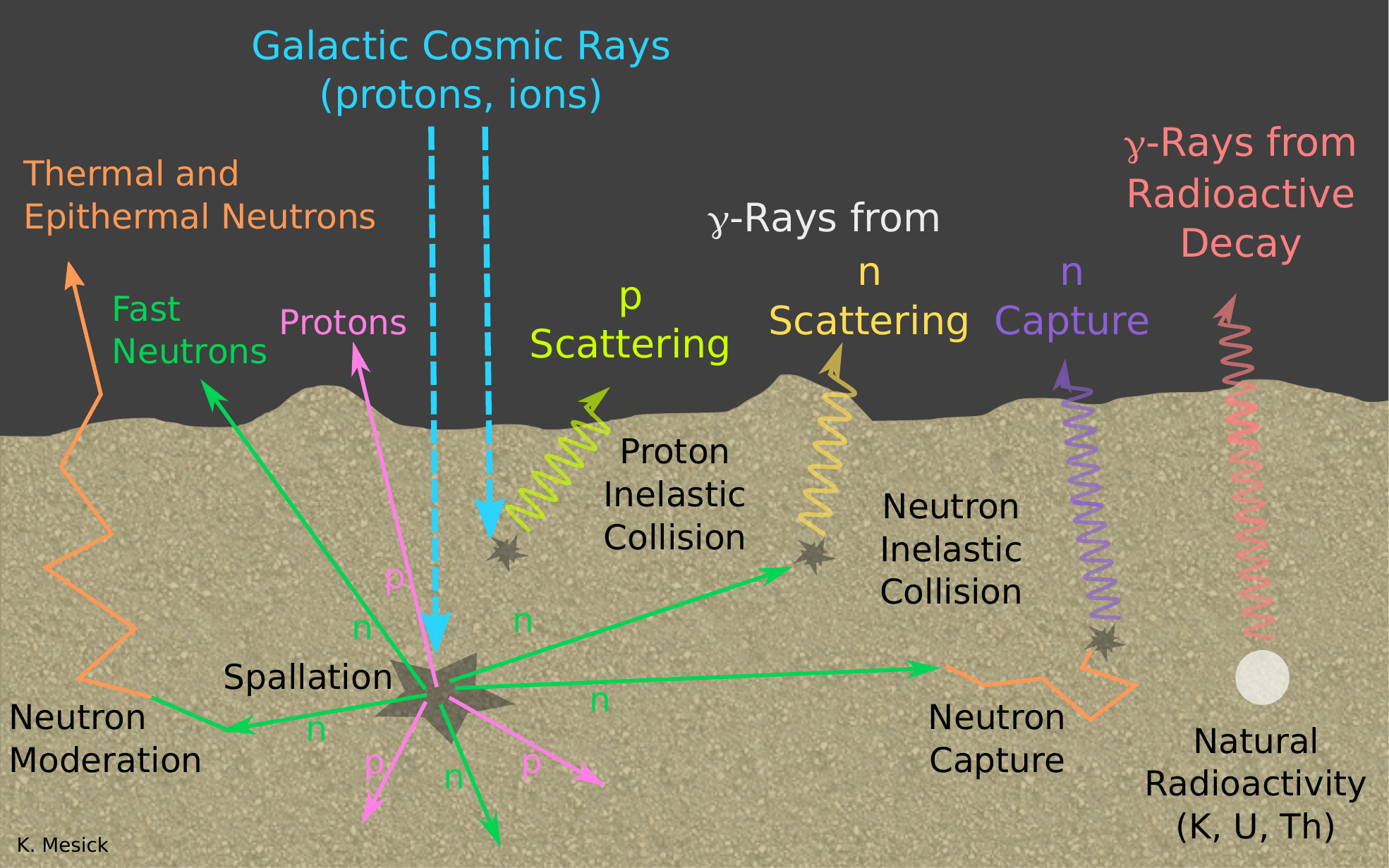}
\caption{Cartoon of GCR spallation in a planetary surface and the signals produced.}
\label{fig:gcrcartoon}
\end{figure}

On planetary bodies with little to no atmosphere, GCRs can hit the body and produce fast neutrons \added{and protons}\deleted{ primarily }through nuclear spallation within the top few meters of the surface.  As illustrated in \explain{Figure 1 updated}\replaced{Fig}{Fig}.~\ref{fig:gcrcartoon}, \replaced{these}{the} neutrons are slowed through inelastic and elastic collisions with elements in the planetary surface and some will escape the surface.  This neutron leakage provides a measure of the average atomic mass of the near-surface material and is highly sensitive to the presence of hydrogen which readily thermalizes neutrons.  \added{High-energy protons also escape the surface but at a much smaller rate.}  Gamma-rays can also escape the surface, produced at characteristic energies from neutron inelastic or neutron capture reactions with elements in the planetary surface or through the decay of natural radioactive isotopes present. \added{Direct interactions of GCR protons with nuclei resulting in nuclear de-excitation can produce a small number of gamma rays that also escape the surface.} These leakage signals can be detected by landed or orbiting neutron\added{, proton,} and gamma-ray spectroscopy instruments and provide distinguishing details about the composition of planetary surfaces that can inform understanding about the formation and evolution of the planetary body. 

\added{Of focus in this paper,} neutron and gamma-ray spectroscopy of GCR-induced signals as a remote sensing technique has been utilized on numerous space missions to study a) the elemental composition of the Moon \citep{Lawrence1998,Lawrence2000,Lawrence2002,Prettyman2006}, Mars \citep{Evans2006,Boynton2007}, Mercury \citep{Peplowski2011,Peplowski2012,Peplowski2014,Peplowski2015,Evans2012,Evans2015}, and the asteroids Vesta \citep{Lawrence2012a,Peplowski2013,Yamashita2013,Prettyman2015} and Eros \citep{Evans2001,Peplowski2016} and b) discover subsurface hydrogen on the Moon \citep{Feldman1998,Feldman2001,Mitrofanov2010,Mitrofanov2012}, Mars \citep{Feldman2002,Feldman2004,Boynton2002,Mitrofanov2002,Mitrofanov2004}, Mercury \citep{Lawrence2013}, Vesta \citep{Prettyman2012}, and Ceres \citep{Prettyman2017}.  Such measurements are aided by the use of radiation transport simulations in the interpretation of results.

This paper studies the validity of the radiation transport code GEometry ANd Tracking (Geant4) version 10.04.p01 \citep{Agostinelli2003,Allison2016} for simulating the interaction of galactic cosmic rays with planetary surfaces by benchmarking simulation results against data from the Apollo 17 Lunar Neutron Probe Experiment (LNPE) \citep{Woolum1975}.  Geant4 is an open-source Monte Carlo based radiation transport toolkit that includes a variety of physics model and data options for handling the interactions of particles with matter over a wide range of energy from eV scale to TeV scale.  A framework for performing simulations is provided, requiring users to implement C++ code to define materials and geometry, choose which physics to implement (or choose from several pre-defined factory physics lists), define the particle source, and define sensitive detectors to extract observables of interest.

In addition to Geant4, we performed the same benchmarking simulation study using the radiation transport code Monte-Carlo N Particle$^{\textregistered}$ version 6.2 (MCNP6) \citep{MCNP6.2}.  This is the latest version of the MCNP code, which saw the efforts and capabilities of MCNP5 and MCNPX, the latter having been widely used in the past for simulations of GCR-induced signals from planetary surfaces, merged and the addition of many new features \citep{Goorley2012}.  
MCNP is developed and maintained at Los Alamos National Laboratory (LANL) and provides a compiled simulation package which defines all available options for physics models, data, geometry, and observables.  
The accessible energy range for particle transport is similar to that of Geant4.  The user does not need to write any source code with MCNP6, instead specifying physics, geometry, and desired observables from the available options within an input file with fixed format.

The layout of this paper is as follows: Section~\ref{sec:previous} discusses previous efforts using radiation transport simulations in this topic area, Section~\ref{sec:sim} presents the simulation method including the data which is used for benchmarking, source term models, physics options, and geometry considerations, Section~\ref{sec:results} presents the simulation results and a discussion of broader implications, and finally Section~\ref{sec:conclusions} summarizes the findings of this work.

\section{Previous Work}\label{sec:previous}

Monte Carlo based simulation efforts in studying the production and transport of GCR-induced particles date back to the 1970s, however, typically exploited multiple codes to simulate different energy regimes of the problem and required numerous approximations.  Armstrong \citep{Armstrong1972} used Monte Carlo methods to calculate the nuclear cascade induced by GCR protons on the lunar surface, however used other methods to transport low-energy neutrons and generate photons.  Work in the 1990s to simulate gamma-ray production from Mars \citep{Masarik1996} and Mercury \citep{Bruckner1997} and to investigate cosmogenic nuclide production in meteorites \citep{Masarik1994} and the Moon \citep{Nishiizumi1997} split the simulation effort into two parts; the high-energy nuclear interactions were simulated with the Monte Carlo transport code LAHET \citep{Prael1989,Prael1993} and the low-energy neutron transport ($<15-20$~MeV) was simulated using early versions of MCNP \citep{Briesmeister1993}.  Around the same time, work using the coupled HERMES code system \citep{Cloth1988} was used to simulate gamma-ray fluxes from Mars \citep{Dagge1991} and cosmogenic nuclide production in meteorites \citep{Bhandari1993}.

Benchmarking of these early simulation efforts used the Apollo 17 LNPE measurements (described in more detail in Section~\ref{sec:apollo}), with results from the LAHET and MCNP coupled simulations \citep{Masarik1996} and the results from the HERMES code system \citep{Dagge1991} showing excellent agreement with the LNPE data.  Despite each of these papers citing excellent agreement with the Apollo 17 LNPE data, these two benchmarking efforts used different normalizations for the total integral GCR flux incident on the lunar surface, differing by almost a factor of two.

More recently, \citep{McKinney2006} performed a thorough study of GCR-induced neutron production on the Moon using MCNPX, showing the impacts of lunar surface composition and physics model choice on the agreement between the MCNPX simulation and the LNPE data.  Both of these choices can lead to 20\% differences in the simulated results, and there is good agreement with the LNPE data for certain physics models when the LNPE drill core material is used.  The radiation transport code PHITS has also been benchmarked against the LNPE data \citep{Ota2011}, and shows reasonable agreement.  Geant4 has been used in a calculation of cosmogenic nuclide production in the Moon \citep{Li2017}, however, no extensive benchmarking of Geant4 physics models to validate its use for this type of application has been performed.

\section{Simulation Method}\label{sec:sim}

\subsection{Galactic Cosmic Ray Source}

Galactic cosmic rays are an isotropic background source that originate from outside the solar system.  \added{The unmodulated source is referred to as the local interplanetary spectrum (LIS).}  Within our own solar system, the heliospheric magnetic field strongly influences the energy distribution of GCRs in local interplanetary space, producing a modulated GCR flux that inversely tracks with solar cycle activity.  GCRs span in energy over several orders of magnitude from $\sim$1~MeV/nucleon (MeV/n) to $\sim$1~TeV/n, peaking in the range of hundreds of MeV/n.  The composition of GCRs is generally quoted in literature as being $\sim$89\% protons, $\sim$10\% alpha particles, and $\sim$1\% heavier nuclei.

\subsubsection{Models}\label{sec:gcrmodel}

There are several models that have been developed to describe the GCR flux spectrum at 1 AU.  These models generally parameterize the energy distribution of GCR protons in terms of sun spot number or an interpreted solar modulation parameter, \replaced{both of these affecting}{which describe the energy loss of the GCRs traveling through the heliospheric magnetic field and} the rigidity cutoff of the spectrum (see \textit{e.g.} \citep{Mrigaskshi2012}). \added{Physically, the energy loss of a GCR travelling through the heliospheric magnetic field is the particle charge multiplied by the solar modulation potential ($\phi$), so that $\phi$ can be considered a true physical property of the heliospheric magnetic field. However, only the combined effect of $\phi$ and the unmodulated LIS can be measured directly within the heliosphere, so that GCR models that employ different LIS parameterizations will require different solar modulation values to reproduce the same GCR spectrum.}  Some parameterizations separately describe alpha particle spectra with independent parameters while others approximate the alpha particles by a scaling of the proton spectrum.  In this paper we consider two models, what we call Usoskin \& Vos/Potgieter (U\&VP) and Castagnoli \& Lal (C\&L).

\begin{figure}[h!]
\centering
\includegraphics[width=0.98\linewidth]{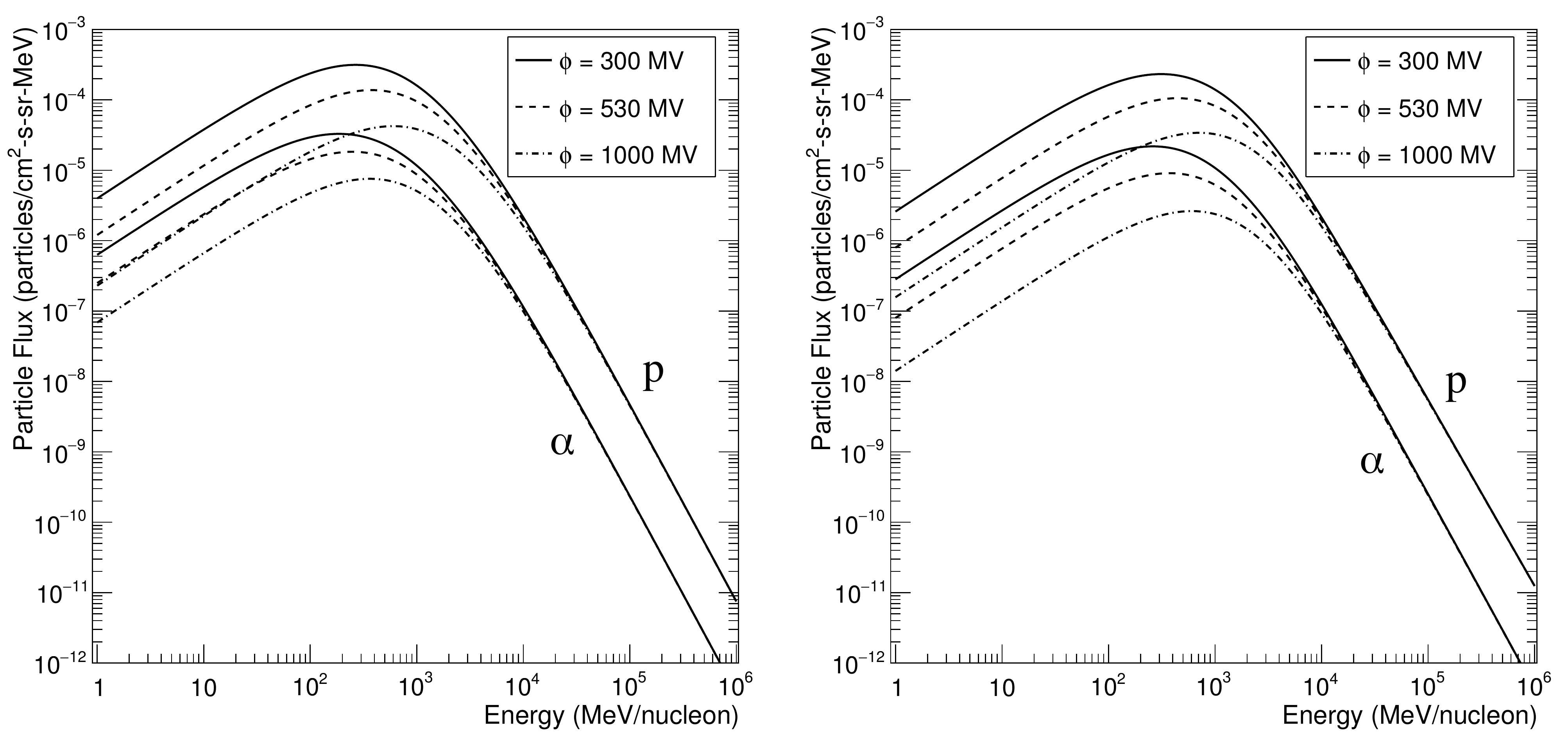}
\caption{GCR differential proton and $\alpha$-particle energy spectra for three values of the solar modulation potential $\phi$ using the Usoskin \& Vos/Potgieter model (left) and Castagnoli \& Lal model (right).}
\label{fig:gcrmod}
\end{figure}

The U\&VP GCR model uses a spherically symmetric force-field approximation \citep{Gleeson1968,Caballero2004} to model the differential energy spectrum.  This parameterization is widely used and takes the form (\textit{e.g.} \citep{Usoskin2005,Usoskin2011,Usoskin2017,Gil2015})
\begin{equation}\label{eq:1}
J_{i}(T) = J_{LIS_{i}}(T+\Phi) \frac{E^2 - E_{r}^2}{(E+\Phi)^2-E_{r}^2}~,
\end{equation}
where for particle species \textsl{GCR}$_i$, $J_{LIS}$ represents \replaced{an unmodulated local interplanetary spectrum (LIS)}{the LIS}, $T$ is the particle's kinetic energy (GeV/n), $E$ is the total energy $T + E_r$ (GeV/n), $E_r$ is the rest mass energy of the proton (0.938 GeV), and $\Phi_i = \phi\left(eZ_i/A_i\right)$ with $Z$ and $A$ being the charge and mass numbers, respectively, and $\phi$ \replaced{an interpreted solar modulation potential (GV) that describes mean energy loss of the GCR particle within the heliosphere}{is the solar modulation potential}.  The resulting $J_i(T)$ is the differential flux of the GCR particle in units of (m$^2$ s sr GeV/n)$^{-1}$.  There are several different choices for the $J_{LIS}$ parameterization.  In early work fitting neutron monitor data to extract solar modulation values by \citep{Usoskin2005,Usoskin2011} and \citep{Gil2015}, a parameterization from \citep{Burger2000} was used.  Other options for $J_{LIS}$ parameterizations are mentioned in the most recent Usoskin \textit{et al.} paper \citep{Usoskin2017}.  For this work we choose the same model selected in \citep{Usoskin2017}, which comes from Vos \& Potgieter \citep{Vos2015}:
\begin{equation}\label{eq:2}
J_{LIS}(T) = 2.7\times10^3 \frac{T^{1.12}}{\beta^2}\left(\frac{T+0.67}{1.67}\right)^{-3.93}~,
\end{equation}
where $T$ is the kinetic energy (GeV/n) and $\beta = v/c$ is the fractional velocity of the proton relative to the speed of light.  This LIS parameterization \added{and five fitting constants} \replaced{was}{were} defined by fitting proton data from PAMELA (Payload for Antimatter Matter Exploration and Light-nuclei Astrophysics) \citep{Vos2015}, so for heavier species this $J_{LIS}$ function is simply scaled, which assumes the protons and heavier GCR species have the same ratio over all energies.  A nucleonic scaling factor of 0.3 is used in \citep{Usoskin2011,Usoskin2017} to account for alpha-particles and heavier species, with the heavier species considered as additional alpha particles.  In our study we choose to use a nucleonic scaling factor of 0.2, consistent with \citep{Usoskin2005}, as we only consider protons and alphas in our simulations.  While heavier GCR particles contribute only $\sim$1\% to the total GCR flux, they will generate more neutrons on average interacting in planetary surfaces than protons and alpha particles.  Ignoring the heavier GCR species means our simulation results may under-estimate measured quantities by up to 10\%.  Note that the LIS spectrum when utilized in Eq.~\ref{eq:1} the kinetic energy $T$ in Eq.~\ref{eq:2} is replaced by ($T + \Phi$), including within $\beta$.  Examples of the differential GCR spectrum for solar modulation values of $\phi = 300$~MV, 530~MV, and 1000~MV as computed in the U\&VP parameterization are shown in Fig.~\ref{fig:gcrmod}.

The C\&L model also uses the \added{spherically symmetric} force-field approximation but with a LIS spectrum following the analytic expression of \citep{Munoz1975}.  The resulting parameterization is defined in \citep{Lal1980,Lal1985} (however, with typos in both papers see \textit{e.g.} \citep{McKinney2006} for the correct equation) and is expressed as:
\begin{equation}\label{eq:3}
g(T,\phi) = A\left(T+ m + \phi\right)^{-\gamma} \frac{T(T + 2E_r)}{(T+\phi)(T+2E_r+\phi)}~,
\end{equation}
where the first term is the LIS and the second term (fraction) is mathematically equivalent to the second term in Eq.~\ref{eq:1}.  Here $g(T, \phi)$ is the differential flux in units of (m$^2$ s sr MeV/n)$^{-1}$, $T$ is the kinetic energy (MeV/n), $E_r$ is the rest energy of the proton (938 MeV), $\phi$ is the solar modulation potential (MV), $m = a\times$exp($-bT)$, and $A$, $a$, $b$, and $\gamma$ are constants.  The \added{four fitting} constants were determined by empirical fits to observed data in the 1960s--1970s \citep{Munoz1975} and are for protons: $A = 9.9\times10^8$, $a = 780$~MeV, $b = 2.4\times10^{-4}$~MeV$^{-1}$, $\gamma = 2.65$, and for alphas: $A = 1.8\times10^8$, $a = 660$~MeV, $b = 1.4\times10^{-4}$~MeV$^{-1}$, $\gamma = 2.77$.  In this model the alpha particles are treated separately to the protons with their own fit constants, compared to the scaling approximation used in the U\&VP parameterization.  Examples of the differential GCR spectrum for solar modulation values of $\phi = 300$~MV, 530~MV, and 1000~MV as computed in the C\&L parameterization are shown in Fig.~\ref{fig:gcrmod}.

A qualitative comparison of the two models (see Fig.~\ref{fig:gcrmod} for a side-by-side comparison) shows that the U\&VP model has a higher differential flux for the same value of $\phi$ than the C\&L model.  \added{Specifically, for $\phi = 530$~MV, the total $4\pi$ proton and alpha GCR fluxes are 3.404 particles/cm$^2$-s and 0.312 particles/cm$^2$-s, respectively, for the U\&VP model and 3.045 particles/cm$^2$-s and 0.243 particles/cm$^2$-s, respectively, for the C\&L model.} This is not unexpected, as $\phi$ is a model-dependent quantity and therefore fitting data for a particular date will not necessarily result in the same interpreted $\phi$ from each model representation (this is true for other models of the LIS as well, \citep[see \textit{e.g.,}][]{Usoskin2005}).  

\subsubsection{Solar Modulation}

The most recent and comprehensive analysis of solar modulation values covering $1951-2016$ can be found in \citep{Usoskin2017}, which improves upon the fitting procedure used in earlier work by the same first-author \citep{Usoskin2005,Usoskin2011}.  Ground-based neutron monitor (NM) data are fit with a convolution of a predicted GCR spectrum and a yield function that accounts for the atmospheric effects on the GCRs and NM detector response \citep{Usoskin2017}.  As discussed in Section~\ref{sec:gcrmodel}, the most recent work of \citep{Usoskin2017} uses Eq.~\ref{eq:1} with an improved LIS from \citep{Vos2015} (Eq.~\ref{eq:2}), but in addition to this an improved model of the NM yield function from \citep{Mishev2013} and a more robust cross-calibration to measured GCR spectra measurements above the atmosphere were used to improve the reconstruction of solar modulation values.

\begin{figure}[h!]
\centering
\includegraphics[width=0.65\linewidth]{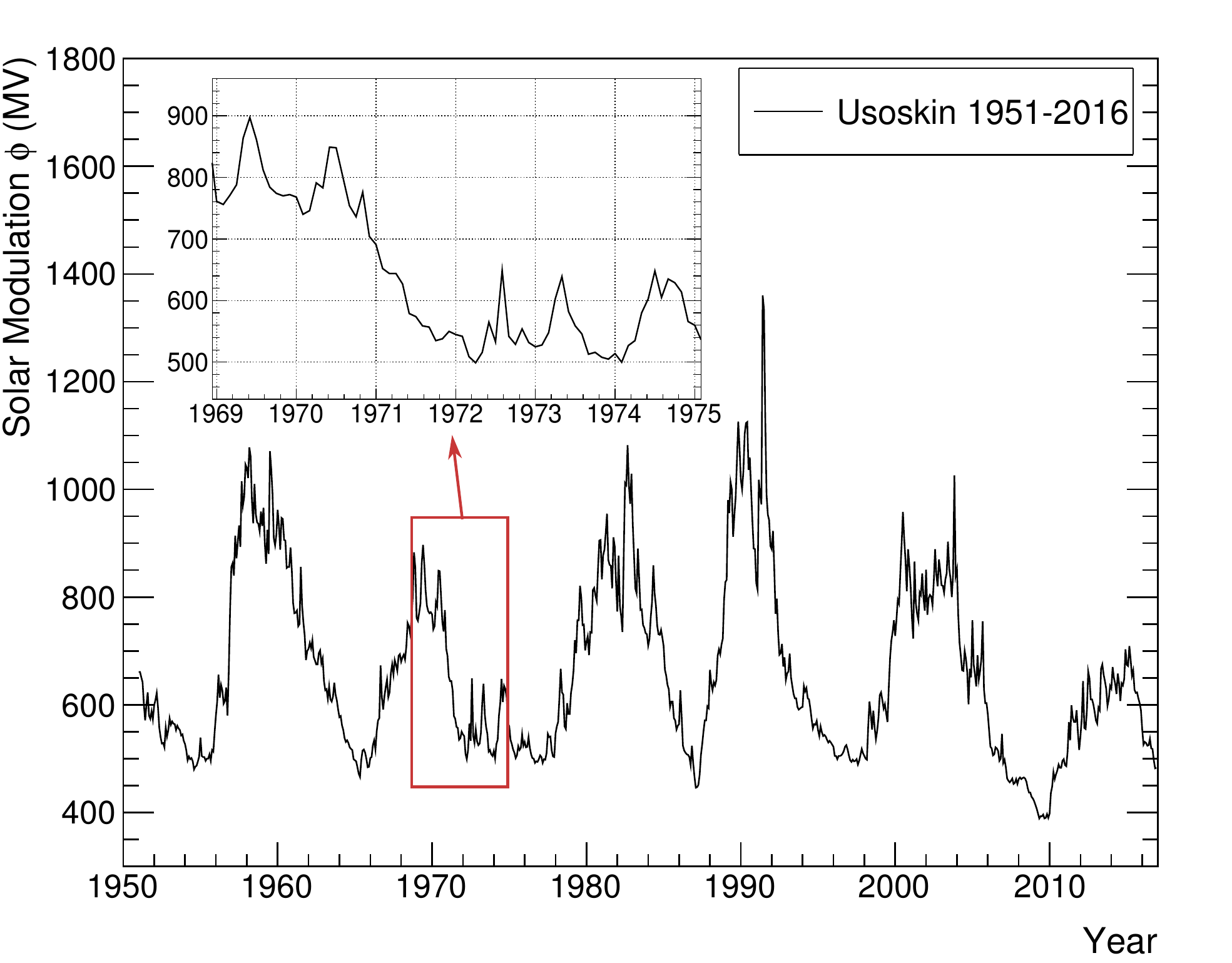}
\caption{Solar modulation values from 1951--2016 from \citep{Usoskin2017}. The red box indicates the region zoomed into in the insert, surrounding the Apollo 17 LNPE measurement.}
\label{fig:smod}
\end{figure}

The interpreted solar modulation values using the U\&VP model in the fitting procedure are presented as monthly averages in Table 2 of \citep{Usoskin2017} and are plotted here in Fig.~\ref{fig:smod}.  Of particular interest in our current work are the value and error in December 1972, when Apollo 17 performed the Lunar Neutron Probe Experiment (LNPE) (see Section~\ref{sec:apollo}).  A region surrounding this event date is indicated in the red box and zoomed into in the figure insert.  From \citep{Usoskin2017} the average solar modulation value for December 1972 is 532~MV with typical errors around this time period of $35-40$~MV.

\subsubsection{Implementation}

In the Geant4 simulations we implemented both GCR parameterizations described above, however, the U\&VP model was used for the primary benchmarking simulations.  The C\&L model was only used in Geant4 to provide a direct comparison to MCNP6, which implements the C\&L parameterization as the default choice for interplanetary analysis.  In Geant4 we used the GeneralParticleSource (GPS) class to input an arbitrary energy histogram for the particle source based on the differential flux given by the U\&VP parameterization for the desired solar modulation values.  In MCNP6, a cosmic source option is available as a particle type, and when a date is specified the solar modulation for that date is interpolated based on the monthly averages determined by \citep{Usoskin2011} \added{through 2011 and from the Usoskin website for 2011--2014 \citep{GreggPC}}.  Since the MCNP6 implementation uses a different GCR parameterization and an older version of Usoskin's solar modulation data, we instead selected arbitrary input dates that resulted in the desired solar modulation values of this study.

For both simulation codes, the GCR energy range was sampled from 1 MeV/n to 1 TeV/n and protons and alpha particles were run separately.  The normalization of GCR source spectrum to total integral flux is calculated automatically in MCNP6 based on the C\&L parameterization if the wgt card is not specified.  In Geant4 the GPS input source spectrum is always renormalized to a unity integral flux, and therefore the output results must be normalized appropriately.  The models prescribed above (Eqs.~\ref{eq:1} and \ref{eq:2} for U\&VP and Eq.~\ref{eq:3} for C\&L) can be multiplied by 4$\pi$ to obtain the omni-directional 4$\pi$ integral flux.  As described in \citep{McKinney2006}, the 4$\pi$ integral flux must then be divided by a factor of 4 to account for the nature of simulating a surface source and only generating events into 2$\pi$.  

To get adequate statistics ($<$1\% errors) we simulated 1 million primary protons and 0.5 million primary alpha particles.  In addition to these particles, Geant4 automatically tracks all secondaries produced except very short-lived species.  In MCNP6 the mode card is used to specify which particles to track, and in addition to the primary particles (protons, alphas) we chose to track neutrons, pions, muons, and the other light ions (deuterons, tritons, and helions).  The cosmic source option in MCNP6 can simulate protons and alphas simultaneously in the appropriate ratio or separately, and we chose to simulate the particles separately.

\subsection{Apollo 17 Data}\label{sec:apollo}

The Apollo 17 mission included the Lunar Neutron Probe Experiment (LNPE) \citep{Woolum1975}, which measured neutron capture rates as a function of depth in the top 2~meters of the Moon's regolith.  A 2-m long detector probe was inserted into a drill core hole for 49 hours and then returned to Earth for analysis.  Of relevance to this paper, a cellulose triacetate plastic detector surrounding 23 $^{10}$B targets recorded alpha particles tracks generated by the neutron capture reaction $^{10}$B(n,$\alpha$)$^{7}$Li.  The neutron densities are reported for 12 depths in \citep{Woolum1975} and shown in Table~\ref{table:lnpe}.  In our comparisons to simulation predictions, we plot these data with errors of 8\%, between the $7-9\%$ error quoted in \citep{Woolum1975} and the same error used in the MCNPX benchmarking of \citep{McKinney2006}.
\begin{table}[h!]
\centering
\caption{Neutron density measured by LNPE \citep{Woolum1975} and interpreted by \citep{McKinney2006}.}
\label{table:lnpe}
\begin{tabular}{cc}
Depth (g/cm$^2$) & Neutron Density (n/cm$^3$) \\
\hline
49 & $5.00\times10^{-6}$ \\
78 & $6.50\times10^{-6}$ \\
113 & $8.65\times10^{-6}$ \\
145 & $9.15\times10^{-6}$ \\
192 & $8.50\times10^{-6}$ \\
236 & $7.80\times10^{-6}$ \\
278 & $6.95\times10^{-6}$ \\
295 & $6.10\times10^{-6}$ \\
325 & $5.35\times10^{-6}$ \\
339 & $5.00\times10^{-6}$ \\
355 & $4.25\times10^{-6}$ \\
383 & $3.80\times10^{-6}$
\end{tabular}
\end{table}

\subsection{Physics Models}

Physics models are often required in radiation transport simulations where data is not readily available.  The choice of physics models can impact the results of a simulation and therefore it is generally a good idea to run multiple models to get an estimate of model uncertainties or to benchmark the models against relevant data.  In the simulation codes we have chosen to run there are two energy regimes for models that are predominantly theory-based, the high-energy regime above a few to tens of GeV up to TeV scale and below this where the intra-nuclear cascade (INC) dominates.  For this application, it is the INC model that is the most important for describing the cascade and the copious secondary particles created by the GCR interactions within the planetary surface.

\subsubsection{Geant4}\label{sec:g4physics}

For the Geant4 simulations we compare five different hadron inelastic physics constructors and three ion physics constructors that are readily available in reference physics lists.  The following physics constructors were implemented for all studies:
\begin{enumerate}
\setlength\itemsep{1pt}
\item \textit{G4HadronElasticPhysicsHP},
\item \textit{G4EmStandardPhysics},
\item \textit{G4DecayPhysics}, and 
\item \textit{G4StoppingPhysics}.
\end{enumerate}
For hadron inelastic physics we considered the constructors:
\begin{enumerate}
\setlength\itemsep{1pt}
\item \textit{G4HadronPhysicsQGSP\_BERT\_HP}, 
\item \textit{G4HadronPhysicsQGSP\_BIC\_HP},
\item \textit{G4HadronPhysicsINCLXX}, 
\item \textit{G4HadronPhysicsFTFP\_BERT\_HP}, and 
\item \textit{G4HadronPhysicsShielding}.  
\end{enumerate}
Table~\ref{table:hadInPhys} shows how the choice of different hadron inelastic physics list constructors affects which model physics and energy ranges are implemented for the nucleons and light mesons.  In the high-energy regime (GeV--TeV) the model options are ``QGSP" and ``FTFP", where ``QGSP" is the Quark-Gluon String Precompound model and ``FTFP" is the FRITOF Precompound model.  Below $\sim$$5-10$~GeV INC models are used, the primary model options being the Bertini cascade, Binary cascade, and Leige INC (INCL \citep{Mancusi2014}).  \added{At the most basic level, the INC models have three steps: a nuclear cascade that results in a highly-excited nucleus, followed by evaporation or fission, then subsequent de-excitation and emission of secondary particles.  In the Bertini model, the production of secondary cascade nucleons is approximated by the interaction of the projectile nucleon with an averaged target nucleus.  Contrary to this, the Binary cascade model treats all individual interactions of primary or secondary cascade particles with individual nucleons of the target nucleus.  The Liege INC model treats the cascade most comprehensively, by tracking the full collision dynamics of all particles within the projectile and target nucleus and composite particles that may be formed.  The models can also differ in the criterion used to signal equilibrium has occurred.}  For neutrons, low-energy ($<20$~MeV) data comes from G4NDL 4.5 (based primarily on ENDF/B-VII.1), and ``HP" refers to the high-precision neutron model.
\begin{table}[h!]
\caption{Energy ranges used in hadron inelastic physics constructors for selected reference physics lists in Geant4 10.4.}
\label{table:hadInPhys}
\smallskip
\begin{adjustbox}{width=1\textwidth}
\centering
\begin{tabular}{|l|c|c|c|c|c|}
\hline
\textbf{Model (protons)} & QGSP\_BERT\_HP & QGSP\_BIC\_HP & QGSP\_INCLXX\_HP & FTFP\_BERT\_HP & Shielding \\
\hline
QGSP & 12 GeV -- 100 TeV & 12 GeV -- 100 TeV & 15 GeV -- 100 TeV & - & - \\
FTFP & 9.5 GeV -- 25 GeV & 9.5 GeV -- 25 GeV & - & 3 GeV -- 100 TeV & 9.5 GeV -- 100 TeV \\
Bertini Cascade & 0 eV -- 9.9 GeV & - & - & 0 eV -- 12 GeV & 0 eV -- 9.9 GeV \\
Binary Cascade & - & 0 eV -- 9.9 GeV & - & - & - \\
Precompond & - & - & 0 eV -- 2 MeV & - & - \\
INCL++v6.0 & - & - & 1 MeV -- 20 GeV & - & - \\
\hline
\textbf{Model (neutron)} & QGSP\_BERT\_HP & QGSP\_BIC\_HP & QGSP\_INCLXX\_HP & FTFP\_BERT\_HP & Shielding \\
\hline
QGSP & 12 GeV -- 100 TeV & 12 GeV -- 100 TeV & 15 GeV -- 100 TeV & - & - \\
FTFP & 9.5 GeV -- 25 GeV & 9.5 GeV -- 25 GeV & - & 3 GeV -- 100 TeV & 9.5 GeV -- 100 TeV \\
Bertini Cascade & 19.9 MeV -- 9.9 GeV & - & - & 19.9 MeV -- 12 GeV & 19.9 MeV -- 9.9 GeV \\
Binary Cascade & - & 19.9 MeV -- 9.9 GeV & - &  - & - \\
INCL++v6.0 & - & - & 19.9 MeV -- 20 GeV & - & - \\
\hline
\textbf{Model (Pions)} & QGSP\_BERT\_HP & QGSP\_BIC\_HP & QGSP\_INCLXX\_HP & FTFP\_BERT\_HP & Shielding \\
\hline
QGSP & 12 GeV -- 100 TeV & 12 GeV -- 100 TeV & 15 GeV -- 100 TeV &- &  - \\
FTFP & 9.5 GeV -- 25 GeV & 4 GeV -- 25 GeV & - & 3 GeV -- 100 TeV & 9.5 GeV -- 100 TeV \\
Bertini Cascade & 0 eV -- 9.9 GeV & 0 eV -- 5 GeV & - & 0 eV -- 12 GeV & 0 eV -- 9.9 GeV \\
INCL++v6.0 & - & - & 0 eV -- 20 GeV & - & - \\
\hline
\textbf{Model (Kaons)} & QGSP\_BERT\_HP & QGSP\_BIC\_HP & QGSP\_INCLXX\_HP & FTFP\_BERT\_HP & Shielding \\
\hline
QGSP & 12 GeV -- 100 TeV & 12 GeV -- 100 TeV & 14 GeV -- 100 TeV & - & - \\
FTFP & 9.5 GeV -- 25 GeV & 4 GeV -- 25 GeV & - & 3 GeV -- 100 TeV & 9.5 GeV -- 100 TeV \\
Bertini Cascade & 0 eV -- 9.9 GeV & 0 eV -- 5 GeV & 0 -- 15 GeV & 0 eV -- 12 GeV & 0 eV -- 9.9 GeV \\
\hline
\end{tabular}
\end{adjustbox}
\end{table}

Table~\ref{table:ionInPhys} shows how the choice of different ion physics list constructors affects which model physics and energy ranges are implemented for ions.  Our three choices of ion physics models are ``Standard" ion physics (\textit{G4IonPhysics}), which calls the binary light-ion cascade model and FTFP, ``QMD" (Quantum Molecular Dynamics - \textit{G4IonQMDPhysics} and \textit{G4IonElasticPhysics}), which also uses the binary light-ion cascade and FTFP models in addition to the QMD model, and ``INCLXX" (\textit{G4IonINCLXXPhysics}), which uses FTFP and the Leige INC.
\begin{table}[h!]
\caption{Energy ranges used in ion inelastic physics constructors for selected reference physics lists in Geant4 10.4.}
\label{table:ionInPhys}
\smallskip
\centering
\begin{adjustbox}{width=1\textwidth}
\begin{tabular}{|l|c|c|c|}
\hline
\textbf{Model (ions)} & Standard & QMD & INCLXX \\
\hline
Binary Light-Ion Cascade & 0 -- 4 GeV/n & 0 - 110 MeV/n & - \\
FTFP & 2 GeV/n -- 100 TeV/n & 9.99 GeV/n -- 1 TeV/n & 2.9 GeV/n -- 1 TeV/n \\
QMD & - & 100 MeV/n -- 10 GeV/n & - \\
INCL++v6.0 & - & - & 0 eV -- 3 GeV/n \\
\hline
\end{tabular}
\end{adjustbox}
\end{table}

The energy ranges used for the physics models (shown in Tables~\ref{table:hadInPhys} and \ref{table:ionInPhys}) are the current defaults in the Geant4 version 10.4 constructors.  When an energy is overlapped by two models, both models are used in a linear combination.  More details can be found in the Geant4 physics reference manual \citep{G4physics}.

\subsubsection{MCNP6}

In MCNP6 the data card \replaced{lca}{\textit{LCA}} can be used to select which physics models are implemented above the energy regime where data are available.  The only high-energy physics model (applicable up to 1~TeV/nucleon) is the Los Alamos Quark-Gluon String Model (LAQGSM) \citep{LAQGSM}.  In the regime of the intra-nuclear cascade there are several model options, with the Cascade-Exciton Model (CEM) \citep{CEM} the default choice.  Other INC options available are the Bertini, ISABEL, or INCL models.  Additionally, the LAQGSM model can be extended into the INC region for light- and heavy-ion interactions.  The combination of CEM + LAQGSM extended to the INC regime is the recommended physics model choice in the MCNP6 manual \citep{MCNP6.2} and produced the best agreement with the LNPE data in the MCNPX benchmarking work by \citep{McKinney2006}.  The data card \replaced{lcb}{\textit{LCB}} can be used to change at what energy or over what energy range the physics models switch between the INC model and the high-energy LAQGSM model.  We used the defaults, where the INC model is used below 3.5~GeV for nucleons and 2.5~GeV for pions while the LAQGSM model is used above these energies.  A range of energies can be specified, and the models will be linearly interpolated between the two energy values.  More details can be found in the MCNP6 reference manual \citep{MCNP6.2}.

\subsection{Geometry Setup}

The geometry was set up similarly in both the Geant4 and MCNP6 simulations.  The Moon was represented as spherical body with a radius of 1738.1~km, with 4 layers of differing material and densities as shown in Table~\ref{table:material}.  As shown and discussed in \citep{McKinney2006}, the choice of lunar composition does have an impact on the simulation results as differing amounts of neutron-absorbing materials, such as titanium and iron, may be present.  The abundance of chlorine, which can greatly impact neutron absorption, is very small (0.75 parts-per-million \citep{Lodders1998}) on the Moon, and therefore negligible.  We follow the work of \citep{McKinney2006} and use the layer compositions from the analysis of the LNPE borehole data.

\begin{table}[h!]
\centering
\caption{Elemental composition of the lunar regolith used in this work, based on the analysis of LNPE borehole data following \citep{McKinney2006}.  Layer depths are from the surface.  Values are in weight percent unless indicated as parts-per-million (ppm).}
\label{table:material}
\begin{adjustbox}{width=0.85\textwidth}
\begin{tabular}{|c|c|c|c|c|}
\hline
LNPE Borehole Analysis & Layer 1 & Layer 2 & Layer 3 & Layer 4 \\
Layer Depth: & $0-22$~cm & $22-71$~cm & $71-224$~cm & $>224$~cm \\
Layer Density: & 1.76~g/cm$^3$ & 2.11~g/cm$^3$ & 1.78~g/cm$^3$ & 1.79~g/cm$^3$ \\
\hline\hline
\multicolumn{5}{|c|}{Elemental (wt\%)} \\
\hline
O  & 0.41739 & 0.41557 & 0.42298 & 0.42636 \\
\hline
Na & 0.00292 & 0.00313 & 0.00307 & 0.00346 \\
\hline
Mg & 0.06162 & 0.06026 & 0.06156 & 0.06091 \\
\hline
Al & 0.06061 & 0.05977 & 0.07384 & 0.07598 \\
\hline
Si & 0.19026 & 0.18955 & 0.19668 & 0.20218 \\
\hline
K (ppm) & 726.24 & 789.65 & 920.11 & 1631.8 \\
\hline
Ca & 0.07541 & 0.07668 & 0.08020 & 0.07707 \\
\hline
Ti & 0.05144 & 0.04905 & 0.03380 & 0.03198 \\
\hline
Cr & 0.00287 & 0.00309 & 0.00264 & 0.00255 \\
\hline
Mn & 0.00176 & 0.00178 & 0.00152 & 0.00146 \\
\hline
Fe & 0.13496 & 0.14030 & 0.12277 & 0.11688 \\
\hline
Sm (ppm) & 8.3342 & 7.7459 & 7.3429 & 10.506 \\
\hline
Eu (ppm) & 1.8164 & 1.7821 & 1.5453 & 1.6997 \\
\hline
Gd (ppm) & 10.997 & 10.577 & 9.8178 & 13.547 \\
\hline
Th (ppm) & 0.9449 & 0.8022 & 1.3819 & 3.0065 \\
\hline
\end{tabular}
\end{adjustbox}
\end{table}

The main simulation observable that we are interested in for comparison with the LNPE data is the neutron density at depth; however, we also calculated the neutron flux at depth and the leakage albedo flux from the surface.  To get the leakage flux (n/cm$^2$-s) we placed a spherical shell 10~cm above the Moon's surface and recorded the\deleted{ current and }flux of particles traveling outward.  In MCNP6 this was achieved with\deleted{ f1 (current) and }\replaced{f2}{\textit{F2}} (flux) tallies, while in Geant4 we recorded energy and angle information for hits crossing the surface by particle type.  To get the neutron density and flux at depth, we created a spherical mesh of 32 layers within the top 300~cm of the surface.  In MCNP6 this was straight forward to implement with the \replaced{SMESH}{\textit{SMESH}} card, which allows specification of a spherical grid that can overlay the geometry.  In Geant4 there is currently no similar option, so instead the planet itself was broken into the appropriate sublayers within the four compositional layers.  The neutron cell flux (n/cm$^2$-s) is the particle weight multiplied by the step length divided by the volume of the cell.  The neutron cell density is the cell flux \replaced{multiplied by the neutron absorption cross section, which is approximated by the (neutron velocity)$^{-1}$}{divided by the neutron velocity}.  In MCNP6 these observables correspond to an \replaced{SMESH}{\textit{SMESH}} tally type ``flux", with the density requiring an \replaced{FM}{\textit{FM}} tally multiplier card to convert the flux to density.  Note that in the version of MCNP6 used, an additional factor of 1$\times$10$^{-8}$ must be used to convert the velocity from units of cm/shake to cm/s.  In Geant4 the neutron flux and neutron density were calculated within sensitive detectors by accessing the required variables (step length, cell volume, and particle energy).

\section{Results \& Discussion}\label{sec:results}

\begin{figure}[h!]
\centering
\includegraphics[width=0.98\linewidth]{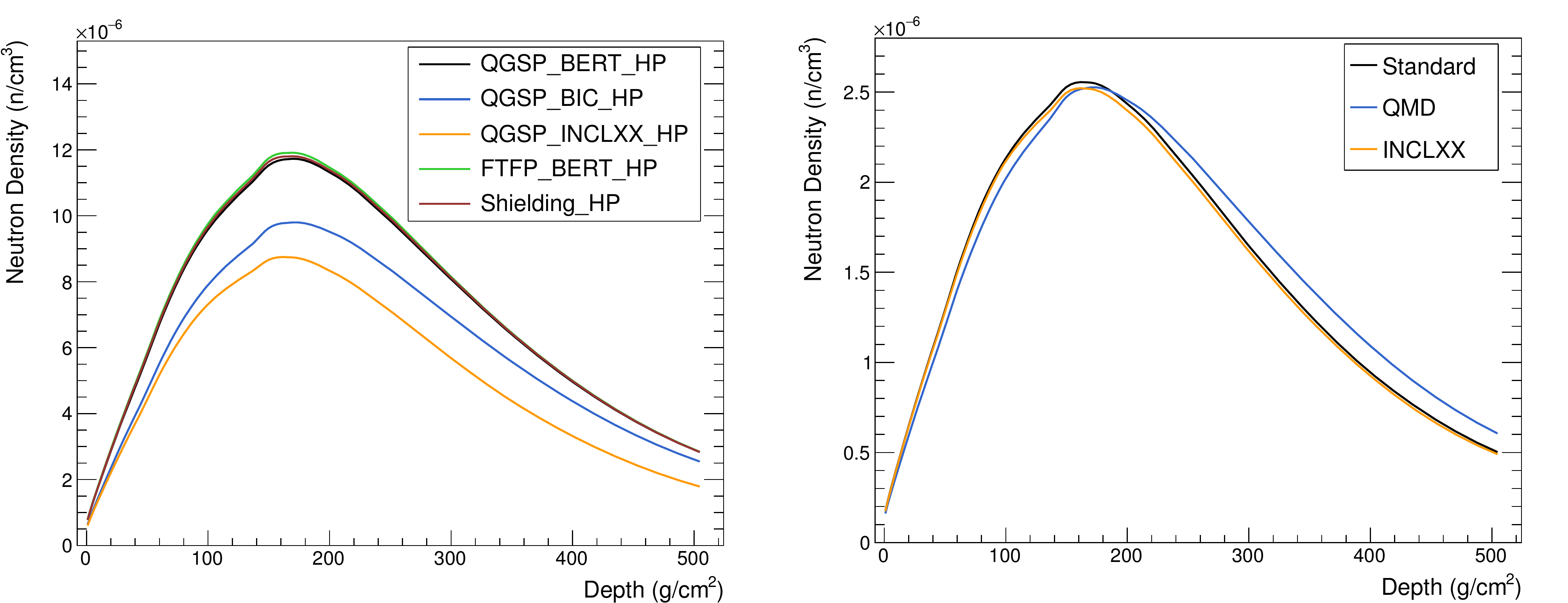}
\caption{Comparison of the Geant4 simulated neutron density for different physics list choices for GCR proton events (left) and GCR alpha events (right) using the U\&VP model with $\phi = 530$~MV.}
\label{fig:g4dens_comp}
\end{figure}

Using the U\&VP model in Geant4, we simulated the neutron density from neutrons with energies between $E = 1\times10^{-10}-1$~MeV for a nominal solar modulation value of $\phi = 530$~MV.  Figure \ref{fig:g4dens_comp} show the neutron density from the Geant4 simulations for GCR proton events and GCR alpha events.  The neutron density resulting from the different choices of physics list constructors described in Section~\ref{sec:g4physics} is compared.  The neutron density from GCR protons is strongly dependent on choice of inelastic hadron physics constructor.  The high-energy physics model (QGSP versus FTFP) does not make a significant difference, as the QGSP\_BERT\_HP and FTFP\_BERT\_HP results are nearly identical.  However, the choice of INC model leads to a range in the simulated neutron density.  The INCL cascade model predicts the smallest neutron density, while the Binary cascade neutron density peak and integral density is $\sim$12\% higher and the Bertini cascade peak and integral density is $\sim$34\% higher.  The lack of significant difference in the simulated neutron density between the FTFP\_BERT\_HP and Shielding\_HP physics lists, which use the same models but applied over different energy ranges, suggests there is little dependence on the selection of model energy ranges within a few GeV.  The neutron density from GCR alphas is not strongly dependent on choice of ion physics constructor.  The Standard and INCL simulated neutron densities are almost identical, while the QMD simulation has a similar shape but is shifted to larger depths.

\begin{figure}[h!]
\centering
\includegraphics[width=0.8\linewidth]{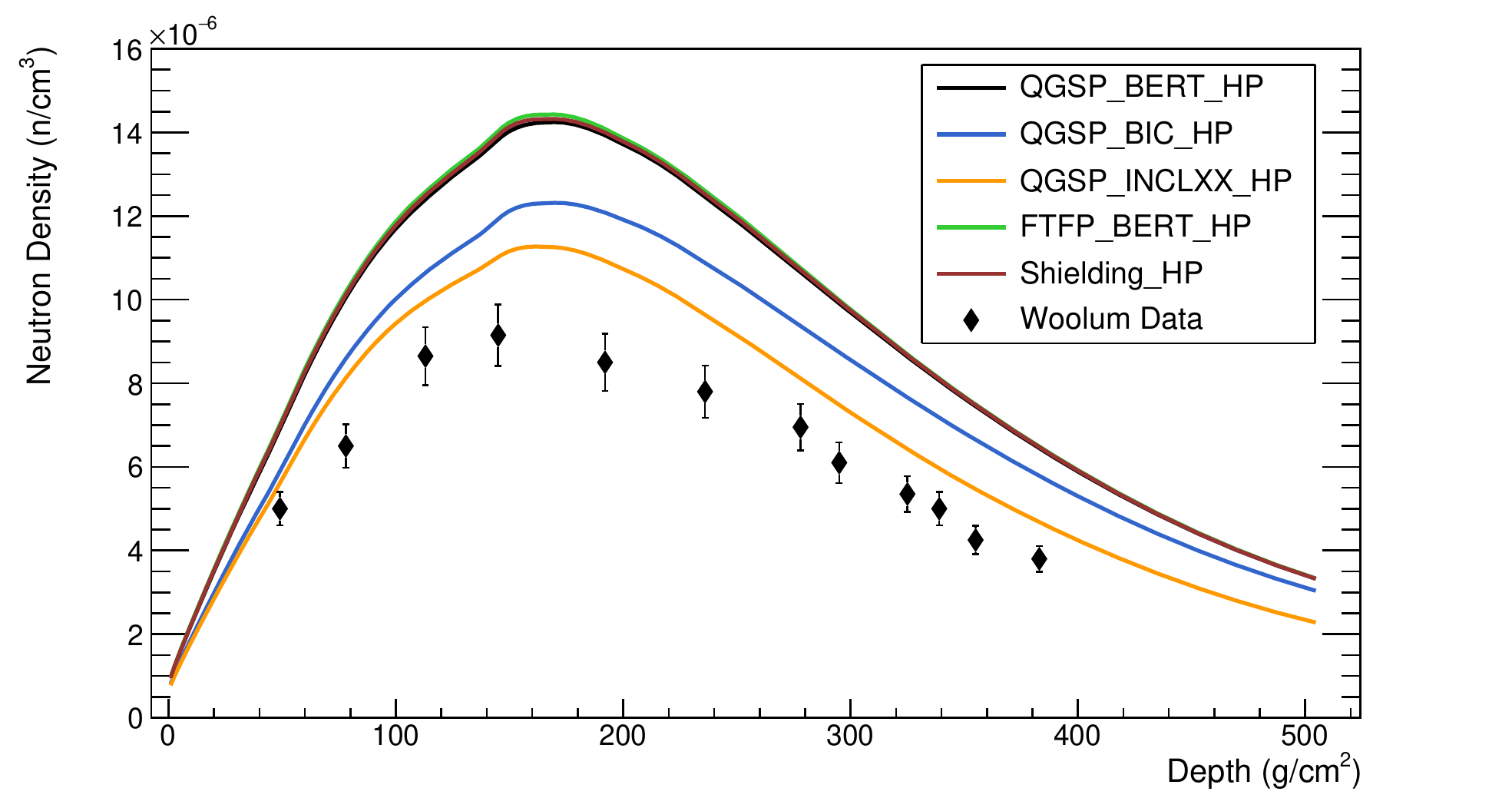}
\caption{Comparison of the Geant4 simulations for total neutron density from GCR protons and alphas with the Woolum LNPE data.  The legend indicates the hadron inelastic physics list used; all curves use the INCLXX physics list for ions.}
\label{fig:g4dens}
\end{figure}

\begin{figure}[h!]
\centering
\includegraphics[width=0.8\linewidth]{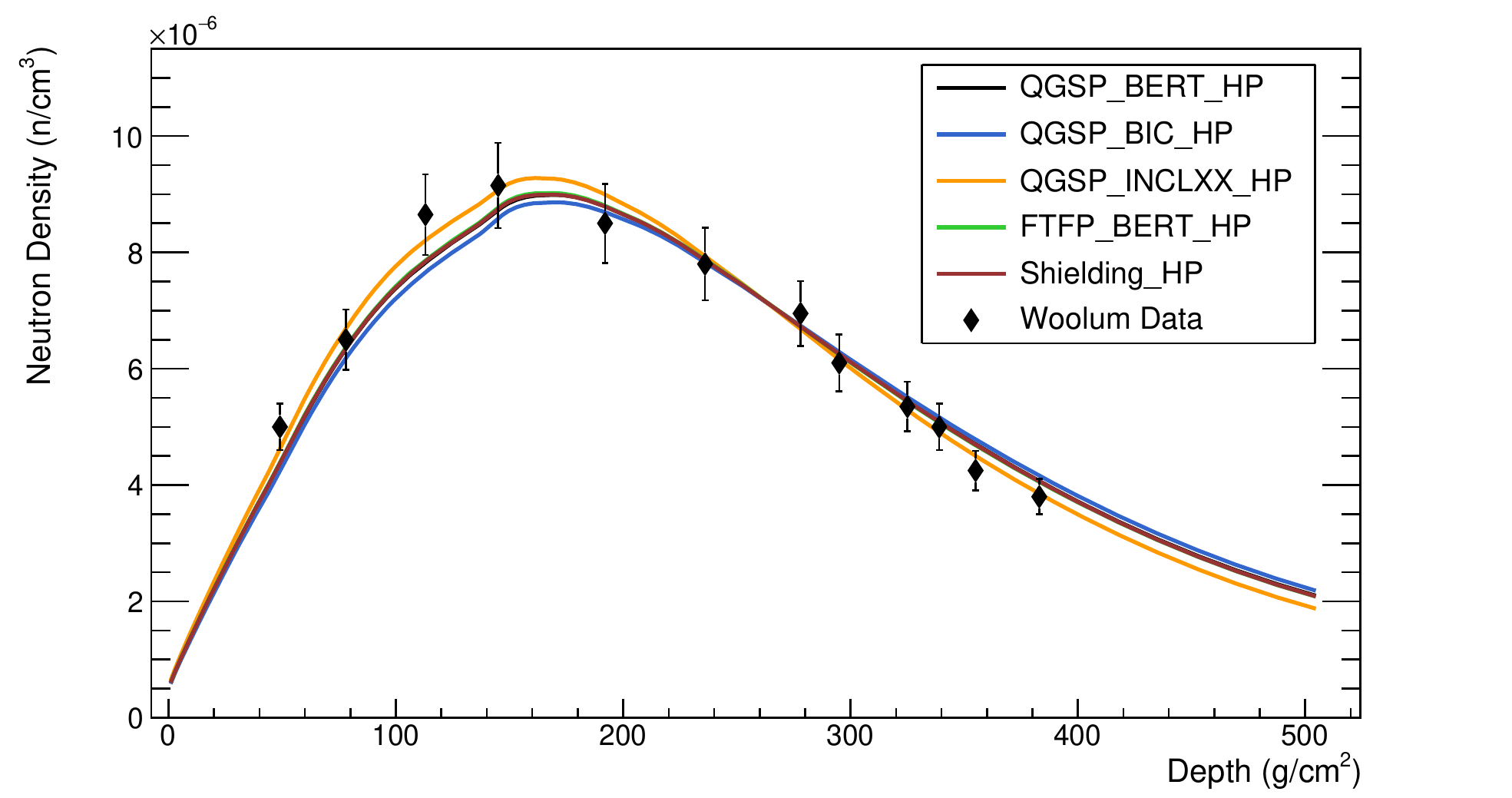}
\caption{Comparison of the scaled Geant4 simulations for total neutron density from GCR protons and alphas with the Woolum LNPE data.  See the text for the scale factors.  The legend indicates the hadron inelastic physics list used; all curves use the INCLXX physics list for ions.}
\label{fig:g4dens_scaled}
\end{figure}

The neutron density from GCR protons for the different physics list choices is added to the GCR alphas as simulated with the INCL ion physics model to get the total neutron density and compared with the Woolum data in Fig.~\ref{fig:g4dens}.  When combined, the GCR alphas contribute about $\sim$20\% of the total neutron density.  All of the Geant4 models we simulated are high relative to the Woolum data.  To see if the shape of these curves matches the data, we determined a multiplicative scale factor for each physics list option based on a $\chi^2$ minimization.  The resulting scaled curves are shown in Fig.~\ref{fig:g4dens_scaled}, where the optimized scale factor is 0.63 for physics lists using the Bertini cascade model (QGSP\_BERT\_HP, FTFP\_BERT\_HP, and Shielding\_HP), 0.72 for the Binary cascade model (QGSP\_BIC\_HP), and 0.84 for the Leige INC model (QGSP\_INCLXX\_HP).  The scaled neutron density curves are all very similar, but the INCL model does show slight differences and has a better scaled fit reduced $\chi^2$ than the other models: 0.3 compared to $0.65-1.0$ for the other models.  The INCL model is also the closest to the data when unscaled.

\begin{figure}[h!]
\centering
\includegraphics[width=0.8\linewidth]{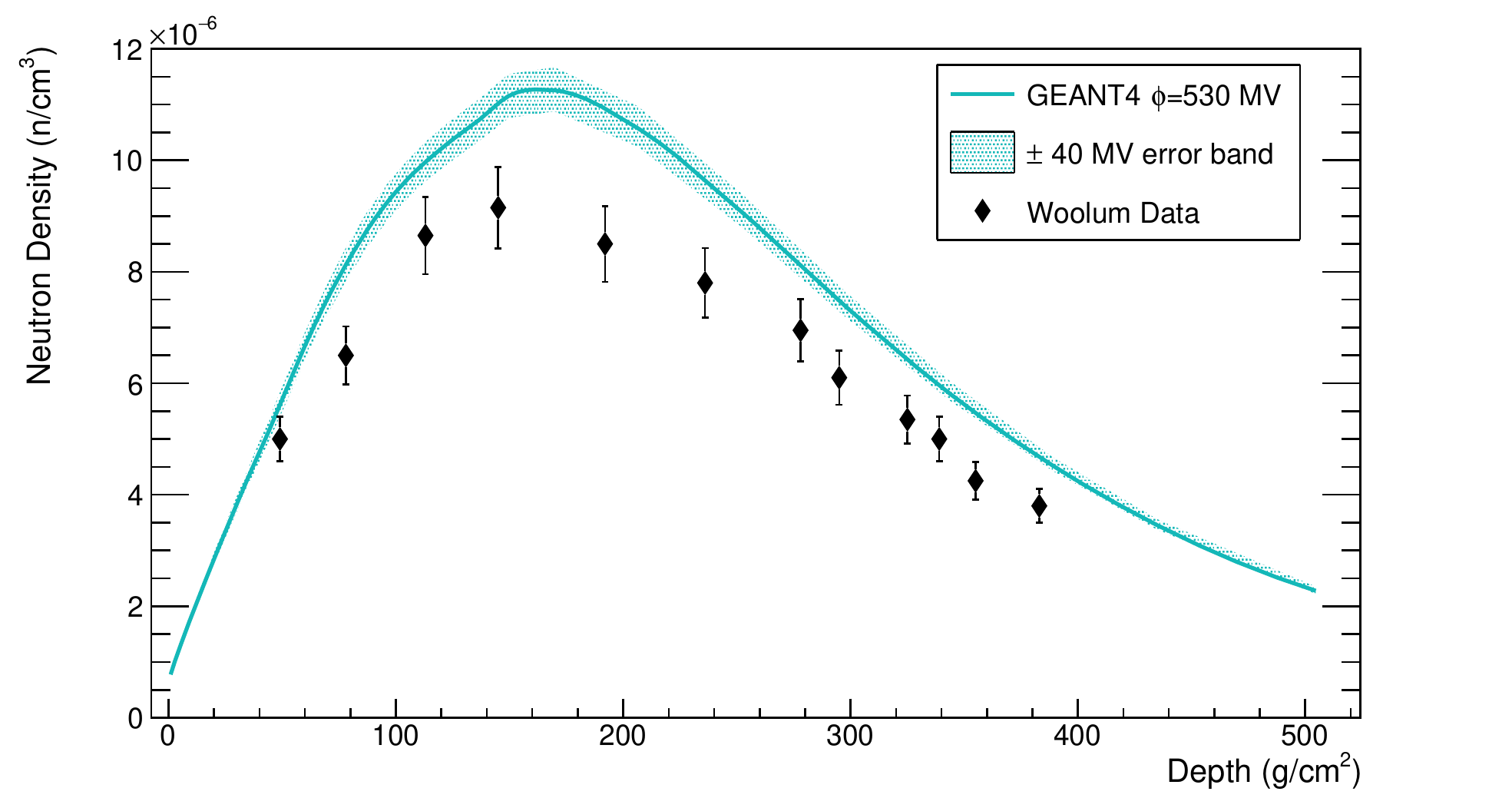}
\caption{The Geant4 simulated neutron density from GCR protons and alphas using the INCL physics models corresponding to the Apollo 17 LNPE measurement and compared with the Woolum data.}
\label{fig:g4densbest}
\end{figure}

This benchmarking case suggests that the INCL model is the best choice for describing the complex INC that is induced by GCR particles hitting planetary surfaces.  This model happens to also be the fastest model in our computations, running 20\% -- 40\% faster than the other models.  As an aside, the Geant4 simulations of cosmogenic nuclide production in the Moon from \citep{Li2017} used the Binary cascade model (QGSP\_BIC\_HP) with modified cross sections to better match data.  Using the physics constructors that use the INCL model (\textit{G4HadronPhysicsINCLXX} and \textit{G4IonINCLXXPhysics}) we also simulated $\phi = 490$~MV and $\phi = 570$~MV, corresponding to $\pm40$~MV uncertainties on the central solar modulation value determined by \citep{Usoskin2017}.  Figure~\ref{fig:g4densbest} compares the neutron density simulated from the nominal value of $\phi = 530$~MV, with error bands determined by the $\pm$40~MV simulations, to the Woolum data.  To match the data without using a scaling factor, we have found a solar modulation of $\phi = 730$~MV is required.

\begin{figure}[h!]
\centering
\includegraphics[width=0.8\linewidth]{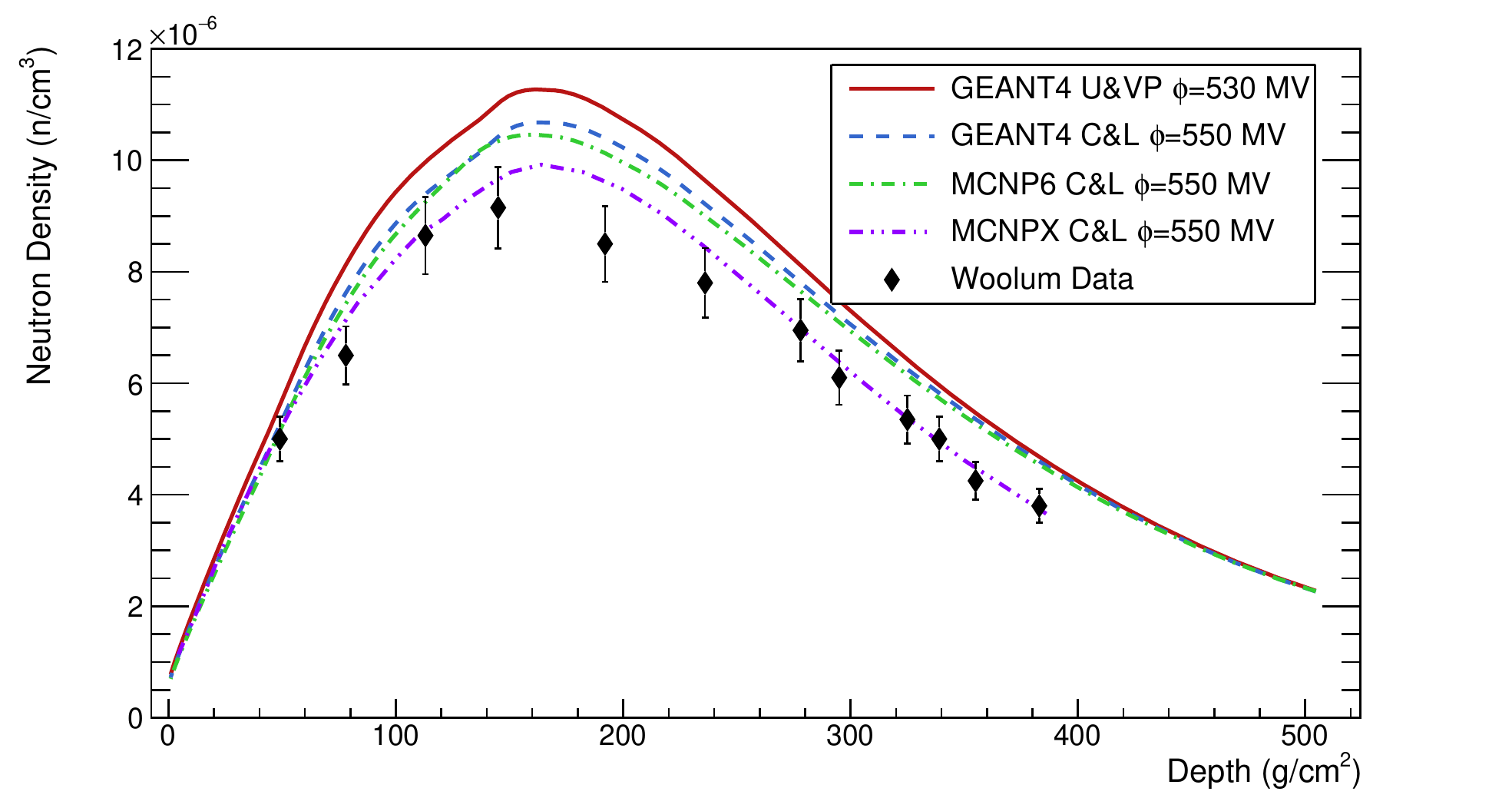}
\caption{The simulated neutron density at depth from GCR protons and alphas for the Geant4 U\&VP model with $\phi=530 $~MV and the C\&L model with \replaced{$\phi=500$~MV}{$\phi=550$~MV} compared with MCNP6 \added{and MCNPX \citep{McKinney2006}}.}
\label{fig:g4mcnp}
\end{figure}

Due to the wide use of MCNPX for this type of simulation in the past, we compare the Geant4 simulated neutron densities using the INCL models to those from MCNP6 using the recommended combination of LAQGSM and CEM physics models.  \explain{Updated Figure 8}\added{To also compare to the MCNPX results from \citep{McKinney2006}, we use the C\&L model in Geant4 with a solar modulation value of $\phi = 550$~MV.}\deleted{  Since the two codes use different GCR parameterizations, we first determined what solar modulation value in the C\&L parameterization implemented in our Geant4 simulation resulted in a neutron density similar to that of $\phi = 530$~MV in the U\&VP model.  This solar modulation value is 500~MV, which is within the errors of the monthly averaged solar modulation in December 1972 from.}  As shown in Fig.~\ref{fig:g4mcnp}, the agreement between Geant4 and MCNP6 for \replaced{$\phi = 500$~MV}{$\phi = 550$~MV} in the C\&L model is fairly good.  The peak of the neutron density is slightly lower in MCNP6, however, the overall shape and scale of the simulation is very similar to the Geant4 result.  The agreement between this version of MCNP6 and the Woolum data is not as good as the benchmarking results using MCNPX \citep{McKinney2006}.  This is likely due to \replaced{changes}{updates} in the physics models and / or normalizations used in the \replaced{simulations}{implementation of the models}.  \added{Note, the use of $\phi = 500$~MV in the C\&L model produces a neutron density similar to that of $\phi = 530$~MV in the U\&VP model.  This solar modulation value is within the errors of the monthly averaged solar modulation in December 1972 from \citep{Usoskin2017}.}

\begin{figure}[h!]
\centering
\includegraphics[width=0.7\linewidth]{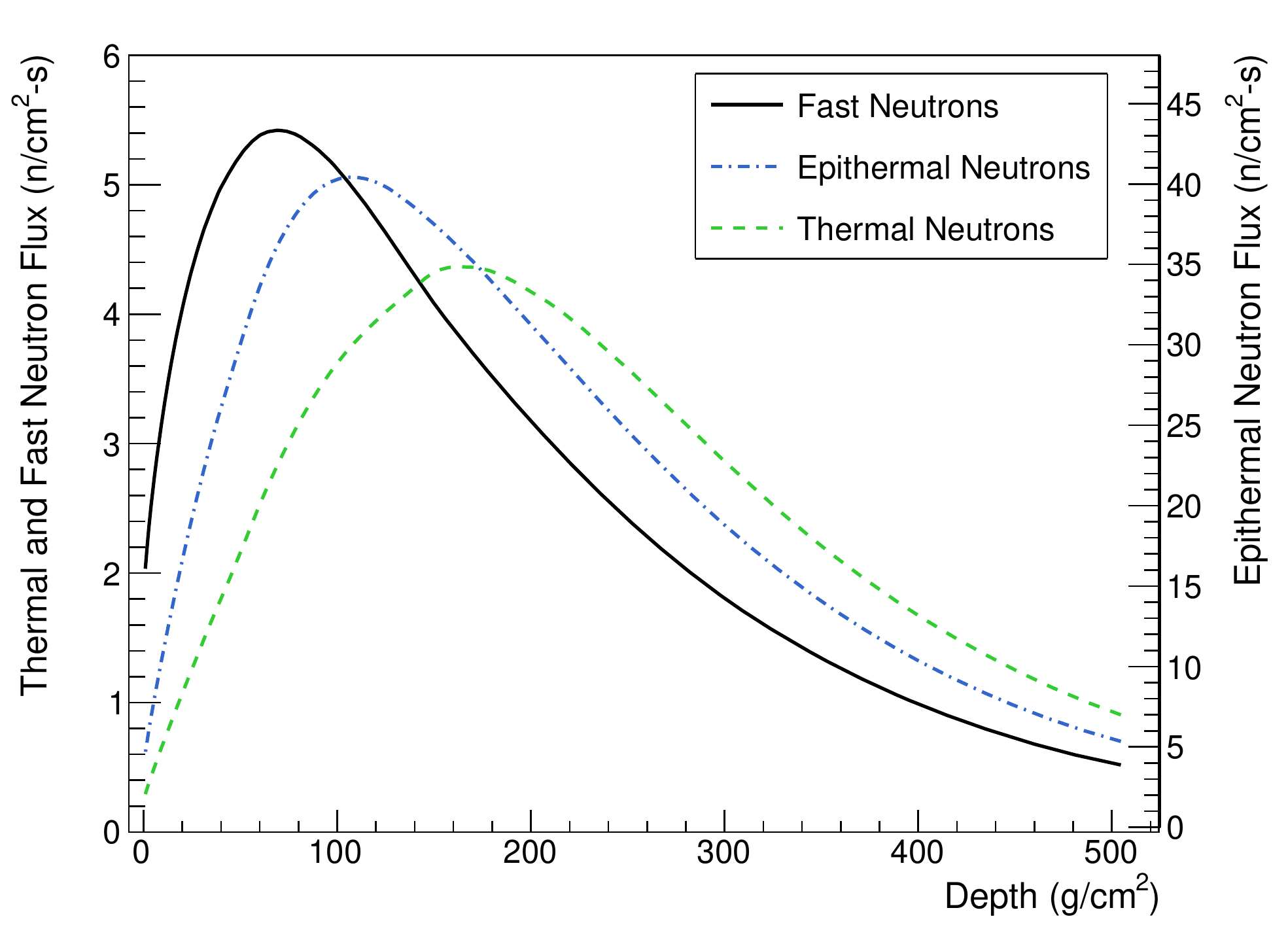}
\caption{The Geant4 simulated neutron flux at depth from GCR protons and alphas for the best physics list.}
\label{fig:g4dflux}
\end{figure}

The neutron flux at depth simulated using the Geant4 INCL models at $\phi = 530$~MV is shown in Fig.~\ref{fig:g4dflux}.  The fast, epithermal, and thermal neutron energy ranges are defined as $E = 1-15$~MeV, $E = 1\times10^{-6}-1$~MeV, and $E = 1\times10^{-10}-1\times10^{-6}$~MeV, respectively.  The fast neutrons peak at more shallow depths than epithermals, which in turn are more shallow than the thermal neutron peak.  The epithermal neutron flux is also much larger than the thermal and fast neutron fluxes (note the two scales in Fig.~\ref{fig:g4dflux}).  These trends are similar to what has been simulated in previous benchmarking to Apollo 17 LNPE data using MCNPX \citep{McKinney2006} and PHITS \citep{Ota2011}.

\begin{figure}[h!]
\centering
\includegraphics[width=0.8\linewidth]{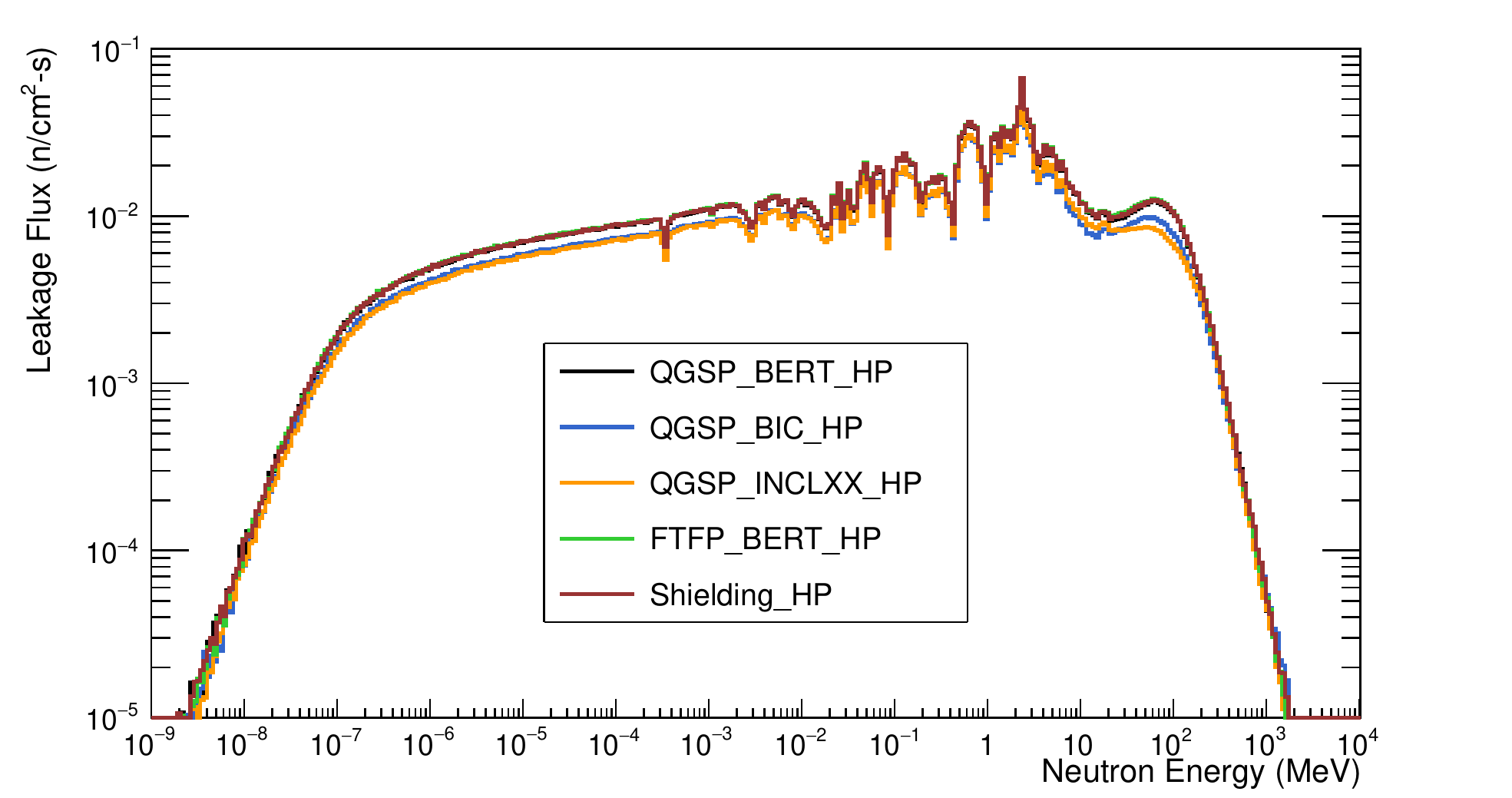}
\caption{The Geant4 simulated neutron leakage \replaced{current}{flux} at 10 cm above the surface of the Moon.}
\label{fig:leakage}
\end{figure}

Finally, a plot comparing the neutron leakage \explain{Updated Figure 10}\replaced{current}{flux} from the Moon, recorded 10 cm above the surface, is shown in Fig.~\ref{fig:leakage}.  In the Geant4 simulations, the Liege INC and Binary cascade models result in a total leakage flux (from $E = 1\times10^{-9}$ MeV -- $1\times10^4$ MeV) of \replaced{3.82}{2.64} n/cm$^{2}$-s, while the physics constructors using the Bertini model are $\sim$\replaced{23}{22}\% higher.  In comparing the Binary cascade and Leige INC model predictions, the thermal and epithermal neutron flux are very similar, while the fast neutron flux shape are very different.  The Binary cascade model fast neutrons look similar in shape to the Bertini model, suggesting the Leige INC model has a different production mechanism for fast neutrons than both the Binary and Bertini cascade models.  However, since the thermal and epithermal neutron flux shapes are similar, this suggests similar scattering and thermalization physics across the models.  The leakage neutron flux from MCNP6 are nearly identical to the results from Geant4 when the INCL model is used with the C\&L model at $\phi = 500$ MV.

\section{Summary}\label{sec:conclusions}

We have simulated lunar neutron density versus depth to compare with data from the Lunar Neutron Probe Experiment.  Simulated results using Geant4 version 10.4 are higher than the data, however, the shape of the neutron density with depth is well reproduced by the simulation.  The multiplicative scale factors depend on which model is used to describe the intra-nuclear cascade physics: the Leige INC simulation is closest to the data with a scale factor of 0.84 and the Bertini cascade model is the farthest from the data with a scale factor of 0.63.  The high-energy physics model does not impact the results.  These scale factors may be off by up to 20\% as we have ignored the contribution from heavier GCR particles, however, the relative differences between the models should remain.  These results suggest that Geant4 is a good tool for helping to interpret trends in neutron and gamma-ray leakage flux measurements, but that an overall normalization must be produced through other means.  The simulated results are likely accurate to within a factor of two.  Results from the MCNP6 simulation are very similar to the Geant4 results using the Leige INC model.

\acknowledgments
The authors thank Gregg McKinney for helpful discussions regarding MCNP \added{and two reviewers for helpful comments}.  Geant4 is an open source code that can be obtained at http://geant4.cern.ch/. \added{Specific details of the Geant4 code developed for this study can be obtained by contacting the corresponding author.} MCNP6 can be requested through the Radiation Safety Information Computational Center at https://rsicc.ornl.gov/Default.aspx.  \added{The MCNP input file used in this study is provided as supplemental material.} Derived data used to generate figures in this manuscript are provided in tabular format as supplemental material.  Research presented in this paper was supported by the Laboratory Directed Research and Development program of Los Alamos National Laboratory under project number 20160672PRD3. Los Alamos National
Laboratory is operated by Los Alamos National Security, LLC under contract to the Department of
Energy’s National Nuclear Security Administration.


\begin{thebibliography}{66}
\providecommand{\natexlab}[1]{#1}
\expandafter\ifx\csname urlstyle\endcsname\relax
  \providecommand{\doi}[1]{doi:\discretionary{}{}{}#1}\else
  \providecommand{\doi}{doi:\discretionary{}{}{}\begingroup
  \urlstyle{rm}\Url}\fi

\bibitem[{\textit{Agostinelli et~al.}(2003)}]{Agostinelli2003}
Agostinelli, S., et~al. (2003), Geant4: A simulation toolkit, \textit{Nuclear
  Instruments and Methods in Physics Research Section A: Accelerators,
  Spectrometers, Detectors and Associated Equipment}, \textit{506}(3), 250 --
  303, \doi{10.1016/S0168-9002(03)01368-8}.

\bibitem[{\textit{Allison et~al.}(2016)}]{Allison2016}
Allison, J., et~al. (2016), {Recent developments in Geant4}, \textit{Nuclear
  Instruments and Methods in Physics Research Section A: Accelerators,
  Spectrometers, Detectors and Associated Equipment}, \textit{835}, 186 -- 225,
  \doi{10.1016/j.nima.2016.06.125}.

\bibitem[{\textit{Armstrong}(1972)}]{Armstrong1972}
Armstrong, T.~W. (1972), Calculation of the lunar photon albedo from galactic
  and solar proton bombardment, \textit{Journal of Geophysical Research},
  \textit{77}(4), 524--536, \doi{10.1029/JA077i004p00524}.

\bibitem[{\textit{Bhandari et~al.}(1993)}]{Bhandari1993}
Bhandari, N., et~al. (1993), Depth and size dependence of cosmogenic nuclide
  production rates in stony meteoroids, \textit{Geochimica et Cosmochimica
  Acta}, \textit{57}(10), 2361 -- 2375,
  \doi{http://dx.doi.org/10.1016/0016-7037(93)90574-G}.

\bibitem[{\textit{Boynton et~al.}(2002)}]{Boynton2002}
Boynton, W.~V., et~al. (2002), {Distribution of Hydrogen in the Near Surface of
  Mars: Evidence for Subsurface Ice Deposits}, \textit{Science},
  \textit{297}(5578), 81--85, \doi{10.1126/science.1073722}.

\bibitem[{\textit{Boynton et~al.}(2007)}]{Boynton2007}
Boynton, W.~V., et~al. (2007), {Concentration of H, Si, Cl, K, Fe, and Th in
  the low- and mid-latitude regions of Mars}, \textit{Journal of Geophysical
  Research: Planets}, \textit{112}(E12S99), \doi{10.1029/2007JE002887}.

\bibitem[{\textit{Briesmeister}(1993)}]{Briesmeister1993}
Briesmeister, J.~F. (1993), {MCNP- A general Monte Carlo N-particle transport
  code, version 4A}, \textit{Los Alamos National Laboratory Report LA-12625-M}.

\bibitem[{\textit{{Br\"{u}ckner and Masarik}}(1997)}]{Bruckner1997}
{Br\"{u}ckner, J. and J. Masarik} (1997), {Planetary gamma-ray spectroscopy of
  the surface of Mercury}, \textit{Planetary and Space Science},
  \textit{45}(1), 39 -- 48,
  \doi{http://dx.doi.org/10.1016/S0032-0633(96)00093-1}.
  
\bibitem[{\textit{Burger et~al.}(2000)\textit{Burger, Potgieter, and
  Heber}}]{Burger2000}
Burger, R.~A., M.~S. Potgieter, and B.~Heber (2000), {Rigidity dependence of
  cosmic ray proton latitudinal gradients measured by the Ulysses spacecraft:
  Implications for the diffusion tensor}, \textit{Journal of Geophysical
  Research: Space Physics}, \textit{105}(A12), 27,447--27,455,
  \doi{10.1029/2000JA000153}.

\bibitem[{\textit{Caballero-Lopez and Moraal}(2004)}]{Caballero2004}
Caballero-Lopez, R.~A., and H.~Moraal (2004), Limitations of the force field
  equation to describe cosmic ray modulation, \textit{Journal of Geophysical
  Research: Space Physics}, \textit{109}(A01101), \doi{10.1029/2003JA010098}.

\bibitem[{\textit{Castagnoli and Lal}(1980)}]{Lal1980}
Castagnoli, G., and D.~Lal (1980), {Solar Modulation Effects in Terrestrial
  Production of Carbon-14}, \textit{Radiocarbon}, \textit{22}(2), 133–158,
  \doi{10.1017/S0033822200009413}.

\bibitem[{\textit{Cloth et~al.}(1988)}]{Cloth1988}
Cloth, P., et~al. (1988), {HERMES Users Guide}, \textit{{Kernforscheungsanlage
  J\"{u}lich Report J\"{u}l-2203}}.

\bibitem[{\textit{{Dagge et~al.}}(1991)}]{Dagge1991}
{Dagge, G. et~al.} (1991), {Monte
  Carlo Simulation of Martian Gamma-Ray Spectra Induced by Galactic Cosmic
  Rays}, \textit{Proceedings of Lunar and Planetary Science}, \textit{21}.
 
\bibitem[{\textit{Evans et~al.}(2006)\textit{Evans, Reedy, Starr, Kerry, and
  Boynton}}]{Evans2006}
Evans, L.~G. et~al. (2006),
  {Analysis of gamma ray spectra measured by Mars Odyssey}, \textit{Journal of
  Geophysical Research: Planets}, \textit{111}(E03S04),
  \doi{10.1029/2005JE002657}.

\bibitem[{\textit{Evans et~al.}(2001)}]{Evans2001}
Evans, L.~G., et~al. (2001), {Elemental composition from gamma-ray spectroscopy
  of the NEAR-Shoemaker landing site on 433 Eros}, \textit{Meteoritics and
  Planetary Science}, \textit{36}(12), 1639--1660,
  \doi{10.1111/j.1945-5100.2001.tb01854.x}.

\bibitem[{\textit{Evans et~al.}(2012)}]{Evans2012}
Evans, L.~G., et~al. (2012), {Major-element abundances on the surface of
  Mercury: Results from the MESSENGER Gamma-Ray Spectrometer}, \textit{Journal
  of Geophysical Research: Planets}, \textit{117}(E00L07),
  \doi{10.1029/2012JE004178}.

\bibitem[{\textit{Evans et~al.}(2015)}]{Evans2015}
Evans, L.~G., et~al. (2015), Chlorine on the surface of mercury: {MESSENGER}
  gamma-ray measurements and implications for the planet’s formation and
  evolution, \textit{Icarus}, \textit{257}, 417 -- 427,
  \doi{https://doi.org/10.1016/j.icarus.2015.04.039}.

\bibitem[{\textit{Feldman et~al.}(1998)}]{Feldman1998}
Feldman, W.~C., et~al. (1998), {Fluxes of Fast and Epithermal Neutrons from
  Lunar Prospector: Evidence for Water Ice at the Lunar Poles},
  \textit{Science}, \textit{281}(5382), 1496--1500,
  \doi{10.1126/science.281.5382.1496}.

\bibitem[{\textit{Feldman et~al.}(2001)}]{Feldman2001}
Feldman, W.~C., et~al. (2001), {Evidence for water ice near the lunar poles},
  \textit{Journal of Geophysical Research: Planets}, \textit{106}(E10),
  23,231--23,251, \doi{10.1029/2000JE001444}.

\bibitem[{\textit{Feldman et~al.}(2002)}]{Feldman2002}
Feldman, W.~C., et~al. (2002), {Global Distribution of Neutrons from Mars:
  Results from Mars Odyssey}, \textit{Science}, \textit{297}(5578), 75--78,
  \doi{10.1126/science.1073541}.

\bibitem[{\textit{Feldman et~al.}(2004)}]{Feldman2004}
Feldman, W.~C., et~al. (2004), {Global distribution of near-surface hydrogen on
  Mars}, \textit{Journal of Geophysical Research: Planets},
  \textit{109}(E09006), \doi{10.1029/2003JE002160}.

\bibitem[{\textit{Garcia-Munoz et~al.}(1975)\textit{Garcia-Munoz, Mason, and
  Simpson}}]{Munoz1975}
Garcia-Munoz, M., G.~M. Mason, and J.~A. Simpson (1975), {The anomalous He-4
  component in the cosmic-ray spectrum at below approximately 50 MeV per
  nucleon during 1972-1974}, \textit{Astrophysical Journal}, \textit{202}(15),
  265--275, \doi{10.1086/153973}.
  
\bibitem[{\textit{Geant4 Collaboration}(2016)}]{G4physics}
  Geant4 Collaboration (2016), {Physics Reference Manual}, \textit{Version 10.3}.

\bibitem[{\textit{Gil et~al.}(2015)}]{Gil2015}
Gil, A., et~al. (2015), Can we properly model the neutron monitor count rate?,
  \textit{Journal of Geophysical Research: Space Physics}, \textit{120}(9),
  7172--7178, \doi{10.1002/2015JA021654}.

\bibitem[{\textit{Gleeson and Axford}(1968)}]{Gleeson1968}
Gleeson, L.~J., and W.~I. Axford (1968), Solar modulation of galactic cosmic
  rays, \textit{Astrophysical Journal J.}, \textit{154}, 1011--1026.

\bibitem[{\textit{Goorley et~al.}(2012)}]{Goorley2012}
Goorley, T., et~al. (2012), {Initial MCNP6 Release Overview}, \textit{Nuclear
  Technology}, \textit{180}(3), 298--315, \doi{10.13182/NT11-135}.
  
\bibitem[{\textit{Gudima et~al.}(2001)}]{LAQGSM}
  Gudima, K.K, S.~M. Mashnik, and A.~Sierk (2001), {User Manual for the Code LAQGSM},
    \textit{Los Alamos National Laboratory Report LA-UR-01-6804}.

\bibitem[{\textit{Lal}(1985)}]{Lal1985}
Lal, D. (1985), Solar-terrestrial relationships and the earth environment in
  the last millennia, \textit{Proceedings of the International School of
  Physics Enrico Fermi}, \textit{95}, 216--233.

\bibitem[{\textit{Lawrence et~al.}(1998)}]{Lawrence1998}
Lawrence, D.~J., et~al. (1998), {Global Elemental Maps of the Moon: The Lunar
  Prospector Gamma-Ray Spectrometer}, \textit{Science}, \textit{281}(5382),
  1484--1489, \doi{10.1126/science.281.5382.1484}.

\bibitem[{\textit{Lawrence et~al.}(2000)}]{Lawrence2000}
Lawrence, D.~J., et~al. (2000), {Thorium abundances on the lunar surface},
  \textit{Journal of Geophysical Research: Planets}, \textit{105}(E8),
  20,307--20,331, \doi{10.1029/1999JE001177}.

\bibitem[{\textit{Lawrence et~al.}(2002)}]{Lawrence2002}
Lawrence, D.~J., et~al. (2002), {Iron abundances on the lunar surface as
  measured by the Lunar Prospector gamma-ray and neutron spectrometers},
  \textit{Journal of Geophysical Research: Planets}, \textit{107}(E12),
  \doi{10.1029/2001JE001530}.

\bibitem[{\textit{Lawrence et~al.}(2013{\natexlab{a}})}]{Lawrence2012a}
Lawrence, D.~J., et~al. (2013{\natexlab{a}}), {Constraints on Vesta's elemental
  composition: Fast neutron measurements by Dawn's gamma ray and neutron
  detector}, \textit{Meteoritics and Planetary Science}, \textit{48}(11),
  2271--2288, \doi{10.1111/maps.12187}.

\bibitem[{\textit{Lawrence et~al.}(2013{\natexlab{b}})}]{Lawrence2013}
Lawrence, D.~J., et~al. (2013{\natexlab{b}}), {Evidence for Water Ice Near
  Mercury's North Pole from MESSENGER Neutron Spectrometer Measurements},
  \textit{Science}, \textit{339}(6117), 292--296,
  \doi{10.1126/science.1229953}.

\bibitem[{\textit{Li et~al.}(2017)}]{Li2017}
Li, Y., et~al. (2017), {Simulation of the production rates of cosmogenic
  nuclides on the Moon based on Geant4}, \textit{Journal of Geophysical
  Research: Space Physics}, \textit{122}(2), 1473--1486,
  \doi{10.1002/2016JA023308}.
  
\bibitem[{\textit{Lodders and Fegley}(1998)}]{Lodders1998}
  Lodders, K. and B. Fegley (1998), \textit{The planetary scientist's companion},
  Oxford University Press.

\bibitem[{\textit{Mancusi et~al.}(2014)}]{Mancusi2014}
Mancusi, D., et~al. (2014), Extension of the li\`ege intranuclear-cascade model
  to reactions induced by light nuclei, \textit{Physical Review C},
  \textit{90}, 054,602, \doi{10.1103/PhysRevC.90.054602}.

\bibitem[{\textit{Masarik and Reedy}(1994)}]{Masarik1994}
Masarik, J., and R.~C. Reedy (1994), Effects of bulk composition on nuclide
  production processes in meteorites, \textit{Geochimica et Cosmochimica Acta},
  \textit{58}(23), 5307 -- 5317,
  \doi{http://dx.doi.org/10.1016/0016-7037(94)90314-X}.

\bibitem[{\textit{Masarik and Reedy}(1996)}]{Masarik1996}
Masarik, J., and R.~C. Reedy (1996), {Gamma ray production and transport in
  Mars}, \textit{Journal of Geophysical Research}, \textit{101}(E8), 18,891 --
  18,912, \doi{10.1029/96JE01563}.

\bibitem[{\textit{Mashnik and Sierk}(2012)}]{CEM}
Mashnik, S., and A.~Sierk (2012), {CEM03.03 User Manual}, \textit{Los Alamos
  National Laboratory Report LA-UR-12-01364}.

\bibitem[{\textit{McKinney et~al.}(2006)}]{McKinney2006}
McKinney, G.~W., et~al. (2006), {MCNPX benchmark for cosmic ray interactions
  with the Moon}, \textit{Journal of Geophysical Research: Planets},
  \textit{111}(E06004), \doi{10.1029/2005JE002551}.

\bibitem[\textit{McKinney} (2018)]{GreggPC}
McKinney, G.~W. (2018), \textit{Private Communication}.

\bibitem[{\textit{Mishev et~al.}(2013)\textit{Mishev, Usoskin, and
  Kovaltsov}}]{Mishev2013}
Mishev, A.~L., I.~G. Usoskin, and G.~A. Kovaltsov (2013), {Neutron monitor
  yield function: New improved computations}, \textit{Journal of Geophysical
  Research: Space Physics}, \textit{118}(6), 2783--2788,
  \doi{10.1002/jgra.50325}.

\bibitem[{\textit{Mitrofanov et~al.}(2002)}]{Mitrofanov2002}
Mitrofanov, I., et~al. (2002), {Maps of Subsurface Hydrogen from the High
  Energy Neutron Detector, Mars Odyssey}, \textit{Science}, \textit{297}(5578),
  78--81, \doi{10.1126/science.1073616}.

\bibitem[{\textit{Mitrofanov et~al.}(2004)}]{Mitrofanov2004}
Mitrofanov, I.~G., et~al. (2004), {Soil Water Content on Mars as Estimated from
  Neutron Measurements by the HEND Instrument Onboard the 2001 Mars Odyssey
  Spacecraft}, \textit{Solar System Research}, \textit{38}(4), 253--257,
  \doi{10.1023/B:SOLS.0000037461.70809.45}.

\bibitem[{\textit{Mitrofanov et~al.}(2010)}]{Mitrofanov2010}
Mitrofanov, I.~G., et~al. (2010), {Hydrogen Mapping of the Lunar South Pole
  Using the LRO Neutron Detector Experiment LEND}, \textit{Science},
  \textit{330}(6003), 483--486, \doi{10.1126/science.1185696}.

\bibitem[{\textit{Mitrofanov et~al.}(2012)}]{Mitrofanov2012}
Mitrofanov, I.~G., et~al. (2012), {Testing polar spots of water-rich permafrost
  on the Moon: LEND observations onboard LRO}, \textit{Journal of Geophysical
  Research: Planets}, \textit{117}(E12), \doi{10.1029/2011JE003956}.

\bibitem[{\textit{Mrigakshi et~al.}(2012)}]{Mrigaskshi2012}
Mrigakshi, A.~I. et~al. (2012), Assessment of galactic cosmic ray models,
  \textit{Journal of Geophysical Research: Space Physics},
  \textit{117}(A08109), \doi{10.1029/2012JA017611}.

\bibitem[{\textit{Nishiizumi et~al.}(1997)}]{Nishiizumi1997}
Nishiizumi, K., et~al. (1997), {Depth profile of $^{41}$Ca in an Apollo 15
  drill core and the low-energy neutron flux in the Moon}, \textit{Earth and
  Planetary Science Letters}, \textit{148}(3), 545 -- 552,
  \doi{http://dx.doi.org/10.1016/S0012-821X(97)00051-4}.

\bibitem[{\textit{Ota et~al.}(2011)}]{Ota2011}
Ota, S., et~al. (2011), Neutron
  production in the lunar subsurface from alpha particles in galactic cosmic
  rays, \textit{Earth, Planets and Space}, \textit{63}(1), 25--35,
  \doi{10.5047/eps.2010.01.006}.

\bibitem[{\textit{Peplowski}(2016)}]{Peplowski2016}
Peplowski, P.~N. (2016), {The global elemental composition of 433 Eros: First
  results from the NEAR gamma-ray spectrometer orbital dataset},
  \textit{Planetary and Space Science}, \textit{134}, 36 -- 51,
  \doi{https://doi.org/10.1016/j.pss.2016.10.006}.

\bibitem[{\textit{Peplowski et~al.}(2011)}]{Peplowski2011}
Peplowski, P.~N., et~al. (2011), {Radioactive Elements on Mercury's Surface
  from MESSENGER: Implications for the Planet's Formation and Evolution},
  \textit{Science}, \textit{333}(6051), 1850--1852,
  \doi{10.1126/science.1211576}.

\bibitem[{\textit{Peplowski et~al.}(2012)}]{Peplowski2012}
Peplowski, P.~N., et~al. (2012), {Variations in the abundances of potassium and
  thorium on the surface of Mercury: Results from the MESSENGER Gamma-Ray
  Spectrometer}, \textit{Journal of Geophysical Research: Planets},
  \textit{117}(E00L04), \doi{10.1029/2012JE004141}.

\bibitem[{\textit{Peplowski et~al.}(2013)}]{Peplowski2013}
Peplowski, P.~N., et~al. (2013), {Compositional variability on the surface of 4
  Vesta revealed through GRaND measurements of high-energy gamma rays},
  \textit{Meteoritics and Planetary Science}, \textit{48}(11), 2252--2270,
  \doi{10.1111/maps.12176}.

\bibitem[{\textit{Peplowski et~al.}(2014)}]{Peplowski2014}
Peplowski, P.~N., et~al. (2014), Enhanced sodium abundance in mercury’s north
  polar region revealed by the {MESSENGER} gamma-ray spectrometer,
  \textit{Icarus}, \textit{228}, 86 -- 95,
  \doi{https://doi.org/10.1016/j.icarus.2013.09.007}.

\bibitem[{\textit{Peplowski et~al.}(2015)}]{Peplowski2015}
Peplowski, P.~N., et~al. (2015), {Geochemical terranes of Mercury’s northern
  hemisphere as revealed by MESSENGER neutron measurements}, \textit{Icarus},
  \textit{253}, 346 -- 363, \doi{https://doi.org/10.1016/j.icarus.2015.02.002}.

\bibitem[{\textit{Prael}(1993)}]{Prael1993}
Prael, R.~E. (1993), {Model cross-section calculations using LAHET}, in
  \textit{Nuclear Data Evaluation Methodology}, edited by C.~L. Dunford, pp.
  525--534, World Scientific, Singapore.

\bibitem[{\textit{Prael and Lichtenstein}(1989)}]{Prael1989}
Prael, R.~E., and H.~Lichtenstein (1989), {User guide to LCS: The LAHET Code
  System}, \textit{Los Alamos National Laboratory Report LA-UR-893014}.

\bibitem[{\textit{Prettyman et~al.}(2006)}]{Prettyman2006}
Prettyman, T.~H., et~al. (2006), {Elemental composition of the lunar surface:
  Analysis of gamma ray spectroscopy data from Lunar Prospector},
  \textit{Journal of Geophysical Research: Planets}, \textit{111}(E12007),
  \doi{10.1029/2005JE002656}.

\bibitem[{\textit{Prettyman et~al.}(2012)}]{Prettyman2012}
Prettyman, T.~H., et~al. (2012), {Elemental Mapping by Dawn Reveals Exogenic H
  in Vesta's Regolith}, \textit{Science}, \textit{338}(6104), 242--246,
  \doi{10.1126/science.1225354}.

\bibitem[{\textit{Prettyman et~al.}(2015)}]{Prettyman2015}
Prettyman, T.~H., et~al. (2015), {Concentrations of potassium and thorium
  within Vesta’s regolith}, \textit{Icarus}, \textit{259}, 39 -- 52,
  \doi{https://doi.org/10.1016/j.icarus.2015.05.035}.

\bibitem[{\textit{Prettyman et~al.}(2017)}]{Prettyman2017}
Prettyman, T.~H., et~al. (2017), {Extensive water ice within Ceres' aqueously
  altered regolith: Evidence from nuclear spectroscopy}, \textit{Science},
  \textit{355}(6320), 55--59, \doi{10.1126/science.aah6765}.

\bibitem[{\textit{Usoskin et~al.}(2005)}]{Usoskin2005}
Usoskin, I.~G. et~al. (2005),
  {Heliospheric modulation of cosmic rays: Monthly reconstruction for
  1951–2004}, \textit{Journal of Geophysical Research: Space Physics},
  \textit{110}(A12108), \doi{10.1029/2005JA011250}.

\bibitem[{\textit{Usoskin et~al.}(2011)}]{Usoskin2011}
Usoskin, I.~G., G.~A. Bazilevskaya, and G.~A. Kovaltsov (2011), {Solar
  modulation parameter for cosmic rays since 1936 reconstructed from
  ground-based neutron monitors and ionization chambers}, \textit{Journal of
  Geophysical Research: Space Physics}, \textit{116}(A02104),
  \doi{10.1029/2010JA016105}.

\bibitem[{\textit{Usoskin et~al.}(2017)}]{Usoskin2017}
Usoskin, I.~G., A.~Gil, G.~A. Kovaltsov, A.~L. Mishev, and V.~V. Mikhailov
  (2017), {Heliospheric modulation of cosmic rays during the neutron monitor
  era: Calibration using PAMELA data for 2006–2010}, \textit{Journal of
  Geophysical Research: Space Physics}, \textit{122}(4), 3875--3887,
  \doi{10.1002/2016JA023819}.

\bibitem[{\textit{Vos and Potgieter}(2015)}]{Vos2015}
Vos, E.~E., and M.~S. Potgieter (2015), {New Modeling of Galactic Proton
  Modulation during the Minimum of Solar Cycle 23/24}, \textit{The
  Astrophysical Journal}, \textit{815}(2), 119.

\bibitem[{\textit{Werner et~al}(2017)}]{MCNP6.2}
Werner, C.J. editor (2017), {MCNP Users Manual - Code Version 6.2}",
  \textit{Los Alamos National Laboratory Report, LA-UR-17-29981}.

\bibitem[{\textit{Woolum et~al.}(1975)\textit{Woolum, Burnett, Furst, and
  Weiss}}]{Woolum1975}
Woolum, D.~S., D.~S. Burnett, M.~Furst, and J.~R. Weiss (1975), Measurement of
  the lunar neutron density profile, \textit{The moon}, \textit{12}(2),
  231--250, \doi{10.1007/BF00577879}.

\bibitem[{\textit{Yamashita et~al.}(2013)}]{Yamashita2013}
Yamashita, N., et~al. (2013), Distribution of iron on vesta,
  \textit{Meteoritics and Planetary Science}, \textit{48}(11), 2237--2251,
  \doi{10.1111/maps.12139}.

\end{thebibliography}

\end{document}